\documentclass[pre,preprint]{revtex4}

\usepackage{amsmath}  
\usepackage{amsfonts} 
\usepackage{graphicx} 
\usepackage{subfig}
\usepackage{enumitem} 
\usepackage{xcolor} 
\usepackage[utf8]{inputenc} 
\usepackage{tikz} 

\begin{document}


\title{Relaxation time of the global order parameter on multiplex networks: the role of interlayer coupling in Kuramoto oscillators}

\author{Alfonso Allen-Perkins$^{1,2}$,Thiago Albuquerque de Assis$^{1,2}$,Juan Manuel Pastor$^{1}$,Roberto F. S. Andrade$^{2}$}
\email{alfonso.allen@hotmail.com}
\email{thiagoaa@ufba.br}
\email{juanmanuel.pastor@upm.es}
\email{randrade@ufba.br}

\affiliation{$^{1}$Complex System Group, Universidad Polit\'ecnica de Madrid, 28040-Madrid, Spain\\$^{2}$Instituto de F\'{i}sica, Universidade Federal da Bahia, 40210-210, Salvador, Brazil}

\date{\today}

\begin{abstract}
This work considers the timescales associated with the global order parameter and the interlayer synchronization of coupled Kuramoto oscillators on multiplexes. For the two-layer multiplexes with initially high degree of synchronization in each layer, the difference between the average phases in each layer is analyzed from two different perspectives: the spectral analysis and the non-linear Kuramoto model. Both viewpoints confirm that the prior timescales are inversely proportional to the interlayer coupling strength. Thus, increasing the interlayer coupling always shortens the transient regimes of both the global order parameter and the interlayer synchronization. Surprisingly, the analytical results show that the convergence of the global order parameter is faster than the interlayer synchronization, and the latter is generally faster than the global synchronization of the multiplex. The formalism also outlines the effects of frequencies on the difference between the average phases of each layer, and identifies the conditions for an oscillatory behavior. Computer simulations are in fairly good agreement with the analytical findings and reveal that the timescale of the global order parameter is at least half times smaller than timescale of the multiplex.
\end{abstract}

\maketitle

\section{Introduction}

The large number of recent investigations on multilayer networks have contributed to uncover several topological and dynamical aspects of complex systems \cite{Boccaletti14,kivela14,buldyrev2010catastrophic,chung1997spectral,sole13,Parshani2010}. These studies have been motivated by the observation that several such systems can be been divided, in a very natural way, into subsets of components that interact in a different way with the co-participants of the same set as compared to members of other subsets. In this way, each such subset can be represented by a layer of multilayer network. This concept has proven to be broad enough to represent different interaction aspects one same agent, provided it also interact differently with members of other subsets \cite{donges2011investigating,gao2012networks,PhysRevE.89.032804}.

Multiplexes form a particular class of multilayer networks, where each layer is formed by the same number $N$ of nodes. Moreover, a multiplex is formed by agents that are identified as one network node, with its own label, in every multiplex layer \cite{PhysRevE.89.032804,sola2013chaos,Domenico2014}. Because of this, each of these agent's representation is connected to its own representations in all other layers \cite{cardillo1,cardillo2,szell}. The strength of these interactions can be dependent of the agent and of the layers between which the interaction occurs \cite{gallotti1,gallotti2,lotero}.

These properties make multiplexes a suitable representation of actual complex systems, where each agent has multiple purposes and abilities. This is the case, for instance, of economic systems where each agent represents an investor that can trade in different world markets. It can use  the communication flow between markets and different market features expressed by local bylaw restrictions to develop strategies in each market to maximize hedge, risk and profits. Under these circumstances, it is natural to ask how and if cooperation and competition \cite{radicchi,comp,reinares,matamalas,wang1,wang2} favor or not the spread of information and synchronization \cite{sorrentino12,gambuzza15,sevilla15,genio16} among the different layers.

To help understand real-world complex dynamics, several synchronous models with non-identical interacting agents have been introduced for a description of synchronization, starting from the R\"{o}ssler and the Kuramoto model \cite{kuramoto75,kuramoto84} in homogeneous structures. More recently, network science explored similar models on non-homogenous structures \cite{Strogatz03, Manrubia04, Kelly11,gambuzza15}. These dynamic models are sufficiently complex to be non trivial and display a large variety of synchronization patterns. Particularly, the Kuramoto model has the advantage of being sufficiently flexible to be adapted to many different contexts and, at the same time, simple enough to be mathematically tractable \cite{Acebron05}. Most of the research done about the Kuramoto model in complex networks has been summarized in the review of Rodrigues \textit{et al.} \cite{rodrigues16}.

The collective dynamics of several interacting populations of Kuramoto oscillators has been investigated on multilayers \cite{ott2008, barreto2008, anderson2012}. Most of the studies on network synchronization focus on effects of network topology on the dynamics in the stationary regime, or when the asymptotic phase of the synchronization is reached. Other investigations have addressed the question of multiplex diffusion \cite{gomez13,sole13}, and the limits it can be enhanced in comparison to the corresponding spread processes in a single layer. However, once the question of how fast the network synchronizes in the steady state is equally important \cite{rodrigues16}, here we want to focus on the difference between diffusion and synchronization speed in multiplexes.  The two phenomena are certainly related but, as we will discuss in the forthcoming sections, they also present different features in the multiplex topology.

\begin{figure}[h!]
\centering
\includegraphics[width=0.6\textwidth]{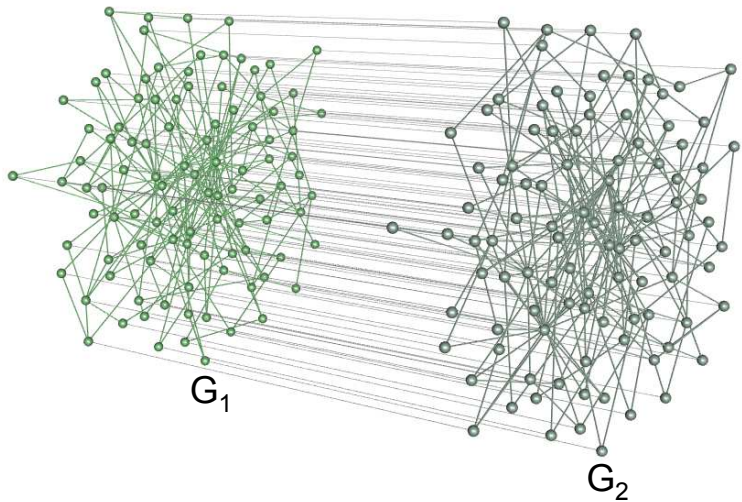}
\caption{Example of an undirected multiplex network with two layers, $G_{1}$ and $G_{2}$ (data visualization with MuxViz \cite{dedomenico15}).}
\label{fig1:ejemplo}
\end{figure}

In this work, we present analytical results for the multiplex order parameter are derived from Kuramoto's equations of motion, both in the linear approximation and in their complete non-linear form, under the assumption that the initial order parameter of each layer is close to unity. Numerical integration of equations of motion corroborate these predictions and present a consistent scenario where it is possible to identify the diffusion relaxation time and the interlayer synchronization phase. As a consequence, the interlayer synchronization is observed to proceed at a non-smaller pace as compared to diffusion.

The paper is organized as follows. In section \ref{Sec:theory}, we define the model and
briefly list the main results of the diffusion relaxation time in multiplexes \cite{gomez13,sole13,Boccaletti14,kivela14,serrano17}. In section \ref{Sec:relax_layers}, the relaxation time of the 
order parameter and of the interlayer synchronization are deduced from spectral analysis and the non-linear Kuramoto model. Numerical results 
supporting the analytical expressions are presented in section \ref{Numerical_results}. Section \ref{Conclusions} summarizes our conclusions.

\section{Kuramoto model in multiplexes and diffusion}
\label{Sec:theory}

We consider initially an undirected multiplex $\mathcal{M}$ with $M$ layers $G_{\alpha}$, $1\leq \alpha \leq M$, where each layer contains $N$ nodes identified by $x_{n}^{\alpha}$, $1\leq n \leq N$ (see  Fig.~\ref{fig1:ejemplo}). A system of coupled Kuramoto oscillators, which takes into account the intra-layer and inter-layer connections, is defined on  $\mathcal{M}$. The oscillator in each node $x_{n}^{\alpha}$ of the layer $G_{\alpha}$ is characterized by its phase $\theta_n^{\alpha}$, whose dynamics is described by

\begin{equation}
\dot{\theta}_{n}^{\alpha}=\Omega _{n}^{\alpha}+\lambda^{\alpha}\sum _{x_{m}^{\alpha} \in G_{\alpha}}w_{nm}^{\alpha} \sin(\theta_{m}^{\alpha}-\theta_{n}^{\alpha})+\mathop{\sum_{ \beta =1}^M}_{\alpha \neq \beta}
\lambda^{\alpha\beta}w_{nn}^{\alpha\beta}\sin(\theta_{n}^{\beta}-\theta_{n}^{\alpha}).
\label{eq:kuramoto}
\end{equation}

\noindent Here, $\Omega _{n}^{\alpha}$ is the natural frequency of the oscillator $x_{n}^{\alpha}$, $\lambda^{\alpha}$ and $\lambda^{\alpha\beta}$  are the coupling strength of the layer $\alpha$ and of the interlayer $\alpha\beta$, respectively, $w_{nm}^{\alpha}$ is  the weight of the connection between the nodes $ x_{n}^{\alpha}$ and $x_{m}^{\alpha}$, and $w_{nn}^{\alpha\beta}$ is the weight of the connection between the nodes $ x_{n}^{\alpha}$ and $x_{n}^{\beta}$. 
In the case of a unweighted and undirected $\mathcal{M}$, $w_{mn}^{\alpha\beta}=1$ and $w_{nm}^{\alpha}=1$ if there is a link between the nodes  $x_{n}^{\alpha}$ and $x_{m}^{\alpha}$, and 0 otherwise.

To present a closer comparison between the results for Eq.~\ref{eq:kuramoto} and those for multiplex diffusion \cite{gomez13,sole13,Boccaletti14,kivela14,KMLee15,dedomenico16}, we consider first the most simple case of undirect $M = 2$ multiplex, without sources and sinks of frequency ($\Omega _{n}^{\alpha}=0$), for which the linear approximation of the Kuramoto model reads

\begin{equation}
\dot{\theta}_n^{\alpha}(t)=\lambda^{\alpha}\sum_{x_{m}^{\alpha} \in G_{\alpha}} w_{nm}^{\alpha}\left ( \theta_m^{\alpha}-\theta_n^{\alpha}
\right )
+
\lambda^{12} \left ( \theta_n^{\beta}-\theta_n^{\alpha}
\right ),
\label{eq:diffusion}
\end{equation}

\noindent with $1 \leq  n,m \leq N$, $1 \leq  \alpha,\beta \leq 2$ and $w_{nn}^{12}=1$.

Once Eq.~\ref{eq:diffusion} is equivalent to the multiplex diffusion equation \cite{gomez13,Boccaletti14}, it can be written as

\begin{equation}
\dot{\vec{\theta}}=-\mathcal{L}\vec{\theta},
\label{eq:laplacian}
\end{equation}

\noindent where $\vec{\theta}$ is a column vector that describes the phase of the oscillators such that $\vec{\theta}^T=\left ( \begin{array}{c|c}
\theta_{1}^{1},\cdots ,\theta_{N}^{1} & \theta_{1}^{2},\cdots ,\theta_{N}^{2}
\end{array} \right )$, $X^T$ stands for the transpose of matrix $X$. $\mathcal{L}$, the \textit{supra-Laplacian matrix} of $\mathcal{M}$, is  defined as

\begin{equation}
 \mathcal{L}=\left ( \begin{array}{c|c}
 \lambda^{1}\mathbf{L}_1 + \lambda^{12}\mathbf{I} & -\lambda^{12}\mathbf{I} \\
\hline
-\lambda^{12}\mathbf{I} &  \lambda^{2}\mathbf{L}_2 + \lambda^{12}\mathbf{I}
\end{array} \right ),
\end{equation}

\noindent where $\mathbf{I}$ is a $N \times N$ identity matrix and $\textbf{L}_{\alpha}$ is the usual $N\times N$ Laplacian matrix of $G_{\alpha}$, with elements $\left ( \mathbf{L}_{\alpha} \right )_{nm}=s_{n}^{\alpha}\delta_{nm} - w_{nm}^{\alpha}$.  $s_{n}^{\alpha}=\sum_{x_{m}^{\alpha}\in G_{\alpha}} w_{nm}^{\alpha}$ and $\delta$ is the Kronecker delta function.

To characterize the eigenvalue spectrum $S(\mathcal{L})\equiv\{\Lambda_i\}$, we rank its eigenvalues in ascending order $0 =  \Lambda_1 <  \Lambda_2 \leq  \dots \leq  \Lambda_{2N}$ \cite{arenas2006a,arenas2006b,gomez13}. The solution of Eq.~\ref{eq:laplacian} in terms of the normal modes $\varphi_i(t)$ is given by
\begin{equation}
\vec{\varphi}=\mathbf{B}^T\vec{\theta},
\end{equation}

\noindent where $\varphi_i(t)= \varphi_i(0)e^{-\Lambda_{i}t}$, and $\mathbf{B}=\left ( \begin{array}{c|c|c|c} \vec{v}_1  & \vec{v}_2 & \dots & \vec{v}_{2N} \end{array} \right )$ is the matrix of eigenvectors of $\mathcal{L}$ (i.e. $\Lambda_i \vec{v}_i=\mathcal{L}\vec{v}_i$) \cite{arenas2006a,arenas2006b,gomez13}.

Consequently, the diffusive relaxation time of multiplex networks, $\tau_{\mathcal{M}}$, depends on the network topology and is dominated by the smallest nonzero eigenvalue $\Lambda_{2}$ of the $\mathcal{L}$, i.e. $\tau_{\mathcal{M}}=1/\Lambda_{2}$ \cite{gomez13,sole13,serrano17}. This behavior is in line with analogous findings for mono-layer networks of coupled Kuramoto oscillators, which have shown that the relaxation time mainly depends on the smallest nonzero eigenvalue of the corresponding Laplacian matrix \cite{Almendral07,Grabow11, Grabow10, Son08}.

If we consider $\lambda^{1}=\lambda ^{2}=1$, the analytical results in \cite{gomez13,sole13} for multiplex diffusion indicate the following properties of $S(\mathcal{L})$:
\begin{enumerate}[label=(\roman*)]
\item $2\lambda^{12}$ is always an eigenvalue of $\mathcal{L}$.
\item When the interlayer coupling is small, i.e. $\lambda^{12} \ll 1$, $\Lambda_{2}=2\lambda^{12}$.
\item When the interlayer coupling is large, i.e. $\lambda^{12} \gg 1$, $\Lambda_{2}\sim \sigma_{s}/2$, where  $\sigma_{s}$ is the smallest nonzero eigenvalue of the superposition matrix $(L_{1}+L_{2})/2$, and $L_{\alpha}$ is the Laplacian matrix of layer $\alpha$.
\end{enumerate}

\begin{figure}[h!]
\centering
\includegraphics[width=0.6\textwidth]{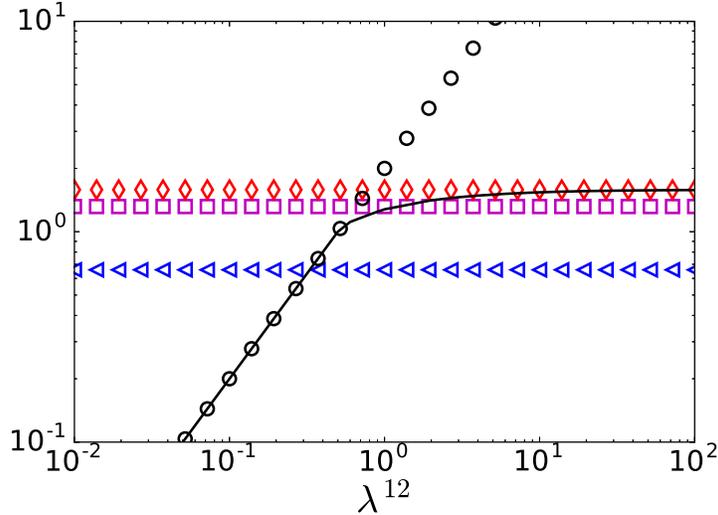}
\caption{Dependence on $\lambda^{12}$ of the second nonzero smallest eigenvalues $\sigma_{2}$ of the Laplacian matrices of layer 1 (blue triangles), layer 2 (magenta squares), the superposition of both layers (red rhombus), $\Lambda_{\Delta}$ 
(black circles) and $\Lambda_{2}$ (black continuous line). The results are presented for a $M=2$ multiplex $\mathcal{M}$ with $N=100$ nodes in each layer, when $\lambda^{1}=\lambda ^{2}=1$. Each layer consists of scale-free network with degree distribution $P(k) \sim k^{-3}$.}
\label{fig:evolucion_Lambda_23}
\end{figure}

In Fig.~\ref{fig:evolucion_Lambda_23} we show an example of the dependence of $\Lambda_{2}$ on $\lambda^{12}$.

\section{Relaxation time of Kuramoto order parameter}
\label{Sec:relax_layers}

The level of synchronization in a general system $\mathcal{S}$ of $\mathcal{N}$ Kuramoto oscillators is described by a parameter $r$  defined as

\begin{equation}
r(t)e^{i\psi(t) }=\frac{1}{\mathcal{N}} \sum _{x_{n}^{\alpha}\in \mathcal{S}}e^{i\theta_{n}^{\alpha}(t) }
\rightarrow
r(t)=\frac{1}{2N}\left | \sum _{x_{n}^{\alpha}\in \mathcal{M}}e^{i\theta_{n}^{\alpha}(t) }\right |,
\label{eq:global_order}
\end{equation}

\noindent where $\psi(t)$ is the average phase of the oscillators in the system. Here, $r\approx 1$ ($r\approx 0$) indicates a full synchronization (an asynchronous behavior) of the system $\mathcal{M}$ \cite{kuramoto75,kuramoto84}.

In this work, Eq.(\ref{eq:global_order}) is used to both layer ($r_\alpha$) and global ($r$) order parameters, by appropriately choosing the set of nodes ($G_\alpha$ or the whole set $\mathcal{M}$) where the sum is performed. $\psi^{\alpha}(t)$ and  $\psi(t)$ indicate $\alpha$-layer and multiplex average phases, respectively. When $M=2$, it is straightforward to express $r$ in terms $r_\alpha$ as

\begin{equation}
re^{i\psi}e^{-i\psi^2}=\frac{r_1e^{i\psi^1}+r_2e^{i\psi^2}}{2}e^{-i\psi^2}\rightarrow r=\sqrt{\frac{r_1^2+r_2^2+2r_1r_2\cos \left ( \psi^1-\psi^2 \right )}{4}}.
\label{eq:global_order_timescale_general}
\end{equation}

For the purpose of putting forward the analytical results, we restrict our analysis to the $r_{\alpha}(t)\approx 1$ case, i.e, we assume that $\theta_{n}^{\alpha}(t) \approx \psi^{\alpha}(t)$ for $1 \leq n \leq N_{\alpha}$, $1\leq \alpha \leq M$, $\forall t$. In section \ref{Numerical_results} we show that these conditions are fairly well satisfied for the system in Eq.~\ref{eq:kuramoto} when, at $t=0$, the degree of synchronization in each layer is high. Under such restrictions, we rewrite $r$ for the $M=2$ case as

\begin{equation}
r(t)\approx \left | \cos \left ( \frac{\psi^1-\psi^2}{2} \right ) \right |= \left | \cos \left ( \frac{\Delta}{2} \right ) \right |,
\label{eq:global_order_timescale}
\end{equation}

\noindent where $\Delta (t) = \psi^{1}(t)-\psi^{2}(t)$ is the difference between the  average phases of the layers $G_1$ and $G_2$. Hence, the timescales of $r$ and $ \left | \cos \left ( \frac{\psi^1-\psi^2}{2} \right ) \right |$ are the same.


The linear relaxation time of the interlayer synchronization process can be estimated by the difference between the average phases of layers $G_1$ and $G_2$, $\Delta$, defined in Eq. \ref{eq:global_order_timescale}. Taking into account the property (i) of $S(\mathcal{L})$, we define  $\Lambda_{\Delta}\equiv2\lambda^{12}$. Its column eigenvector $\vec{v}_{\Delta}$ is such that
$\vec{v}_{\Delta}^T=\left ( \begin{array}{c|c} v_{1}^{1},\cdots ,v_{N}^{1} & v_{1}^{2},\cdots ,v_{N}^{2}
\end{array} \right )=\left ( \begin{array}{c|c}
1,\cdots ,1 & -1,\cdots ,-1
\end{array} \right )$.

By definition $\mathbf{L}_1$ and $\mathbf{L}_2$ are symmetric real matrices with row and column sums zero, i.e. $\mathbf{L}_{\alpha}\vec{1}=\vec{0}$, where $\vec{x}$ is an all-x vector. Thus,

\begin{equation}
 \mathcal{L}\vec{v}_{\Delta}=\left ( \begin{array}{c|c}
 \lambda^{1}\mathbf{L}_1 & \mathbf{0} \\
\hline
\mathbf{0} &  \lambda^{2}\mathbf{L}_2
\end{array} \right )\vec{v}_{\Delta}
+
\left ( \begin{array}{c|c}
 \lambda^{12}\mathbf{I} & -\lambda^{12}\mathbf{I} \\
\hline
-\lambda^{12}\mathbf{I} &  \lambda^{12}\mathbf{I}
\end{array} \right )\vec{v}_{\Delta}
=\vec{0}+2\lambda^{12}\vec{v}_{\Delta}=\Lambda_{\Delta}\vec{v}_{\Delta}.
\end{equation}

Following \cite{arenas2006a,arenas2006b,gomez13}, the normal mode related to $\Lambda_{\Delta}=2\lambda^{12}$ is

\begin{equation}
\vec{v}_{\Delta}^T \vec{\theta}=\sum_{x_n^{1} \in G_{1}} {\theta_n^1}-\sum_{x_m^{2} \in G_{2}} {\theta_m^2}=\varphi_{\Delta}(0)e^{-\Lambda_{\Delta}t}
\label{eq:normal_mode}.
\end{equation}

According to  Eq.~\ref{eq:global_order_timescale}, when the assumption $r_{\alpha}(t)\approx 1$ is valid, Eq.~\ref{eq:normal_mode} leads to

\begin{equation}
\Delta(t)=\psi^{1}(t)-\psi^{2}(t)\approx\frac{\varphi_{\Delta}(0)}{N}e^{-\Lambda_{\Delta}t}.
\label{eq:psi_normal_mode}
\end{equation}

Since the relaxation time for interlayer synchronization can be estimated by $\tau_{\Delta}=1/\Lambda_{\Delta}$, we draw the following similar conclusions to the results listed in section \ref{Sec:theory}:
\begin{enumerate}[label=(\roman*)]
\item When $\lambda^{12} \ll 1$, the diffusive timescale of $\mathcal{M}$ coincides with the interlayer synchronization time, i.e. $\Lambda_{2}=\Lambda_{\Delta}$.
\item When $\lambda^{12} \gg 1$, the diffusive timescale of $\mathcal{M}$ exceeds the interlayer synchronization time, i.e. $\Lambda_{2} \ll \Lambda_{\Delta}$ ($\Leftrightarrow \tau_{\mathcal{M}} \gg \tau_{\Delta}$)  .
\end{enumerate}


To derive the non-linear relaxation timescale of the interlayer synchronization for the system in Eq.~\ref{eq:kuramoto}, we rewrite it in terms of the order parameters $r_\alpha$ of each layer $G_{\alpha}$ as

\begin{equation}
\dot{\theta}_{n}^{\alpha}=\Omega _{n}^{\alpha}+\lambda^{\alpha}r_{\alpha}N\bar{w}_{n}^{\alpha}\sin(\psi^{\alpha}-\theta_{n}^{\alpha})+\mathop{\sum_{ \beta =1}^M}_{\alpha \neq \beta}
\lambda^{\alpha\beta}w_{nn}^{\alpha\beta}\sin(\theta_n^{\beta}-\theta_{n}^{\alpha}),
\label{eq:kuramoto_orden}
\end{equation}

\noindent where $\bar{w}_{n}^{\alpha}$ is defined by

\begin{equation}
\bar{w}_{n}^{\alpha}\sum _{x_{m}^{\alpha}\in G_{\alpha}}e^{i\theta_{m}^{\alpha} }=\sum _{x_{m}^{\alpha}\in G_{\alpha}}w_{nm}^{\alpha}e^{i\theta_{m}^{\alpha} }.
\end{equation}

As $r_{\alpha}(t)\approx 1$, we obtain the following approximation for an undirected multiplex $\mathcal{M}$:

\begin{eqnarray}
\dot{\psi}^{\alpha}=\frac{1}{N} \sum _{x_{n}^{\alpha}\in G_{\alpha}}\dot{\theta}_{n}^{\alpha}=\frac{1}{N}\left [ \sum_{n=1}^{N}\Omega _{n}^{\alpha}\right ]+\mathop{\sum_{ \beta =1}^M}_{\alpha \neq \beta}\lambda^{\alpha\beta}\sin(\psi^{\beta}-\psi^{\alpha})\left [
\sum_{n=1}^{N} w_{nn}^{\alpha\beta}\right ]= \nonumber\\
=\left \langle\Omega\right \rangle_{\alpha}+\mathop{\sum_{ \beta =1}^M}_{\alpha \neq \beta}\lambda^{\alpha\beta}\sin(\psi^{\beta}-\psi^{\alpha})\frac{s^{\alpha\beta}}{N},
\end{eqnarray}

\noindent where $s^{\alpha\beta}$ is the sum of the interlayer strengths between nodes of the layers $G_{\alpha}$ and $G_{\beta}$. Also, the evolution of the average phase difference between  $G_{\alpha}$ and $G_{\beta}$ becomes

\begin{eqnarray}
\dot{\Delta}^{\alpha\beta}=\dot{\psi}^{\alpha}-\dot{\psi}^{\beta}=
\left (\left \langle\Omega\right \rangle_{\alpha}-\left \langle\Omega\right \rangle_{\beta}\right )  -  2\lambda^{\alpha\beta}\sin\left ( \psi^{\alpha}-\psi^{\beta}\right )\frac{s^{\alpha\beta}}{N} \nonumber\\
+\mathop{\sum_{ \gamma =1}^M}_{\gamma \neq \alpha,\beta}\left [
\lambda^{\alpha\gamma}\sin(\psi^{\gamma}-\psi^{\alpha})\frac{s^{\alpha\gamma}}{N}
-\lambda^{\beta\gamma}\sin(\psi^{\gamma}-\psi^{\beta})\frac{s^{\beta\gamma}}{N}
\right ].
\label{eq:general}
\end{eqnarray}

Restricting the discussion to $M = 2$ and $w_{nn}^{12}=1 \Rightarrow s^{12}=N$, we consider first  $\left \langle\Omega\right \rangle_{1}\approx\left \langle\Omega\right \rangle_{2}$, so that the synchronization of the system can be estimated as

\begin{equation}
\eta_{\Delta}(t) \equiv \left | \tan\left ( \frac{\Delta(t)}{2} \right ) \right |=\left | \tan\left ( \frac{\Delta(0)}{2} \right ) \right |e^{-
\int_{0}^{t}2\lambda^{12}
\mathrm{d} {t}'}=
\left | \tan\left ( \frac{\Delta(0)}{2} \right ) \right |e^{-\Lambda_{\Delta}t},
\label{eq:general_approx}
\end{equation}

\noindent where we use the short-hand notation $\Delta(t)=\Delta^{12}(t)$. Eq.~\ref{eq:general_approx} and the series expansion $\tan(x)\simeq x$ show that the relaxation time of $\Delta$ is dominated by $\Lambda_{\Delta}$, i.e., $\Delta/2 \propto  e^{-\Lambda_{\Delta}t}$.

Next, if $\left \langle\Omega\right \rangle_{1}\neq\left \langle\Omega\right \rangle_{2}$, it is possible to integrate  Eq.~\ref{eq:general} and express the corresponding solution in terms of a variable $\xi(t)$ such that

\begin{equation}
\xi(t)=\frac{\left | \tan\left ( \frac{\Delta(t)}{2} \right ) -\mathrm{sgn}\left (\left \langle\Omega\right \rangle^{12}   \right ) \left ( \left | R \right |-\sqrt{R^2-1} \right ) \right |}{ \left | \tan\left ( \frac{\Delta(t)}{2} \right ) -\mathrm{sgn}\left (\left \langle\Omega\right \rangle^{12}   \right )\left ( \left | R \right |+\sqrt{R^2-1} \right ) \right |}=\xi(0)e^{-t\left |\left \langle\Omega\right \rangle^{12}  \right |\sqrt{R^2-1}},
\label{eq:super_general_approx}
\end{equation}

\noindent where $\mathrm{sgn}(.)$ is the sign function, $\left \langle\Omega\right \rangle^{12}\equiv\left \langle\Omega\right \rangle_{1}-\left \langle\Omega\right \rangle_{2}$ and

\begin{equation}
R=\frac{\Lambda_{\Delta}}{\left \langle\Omega\right \rangle_{1}-\left \langle\Omega\right \rangle_{2}}\equiv\frac{\Lambda_{\Delta}}{\left \langle\Omega\right \rangle^{12}}.
\label{eq_R}
\end{equation}

Eq.~\ref{eq:super_general_approx} is valid when $\left | R \right |>1$ while, for the $\left | R \right |\leq1$, the integration of Eq.~\ref{eq:general} results in

\begin{equation}
\tan\left ( \frac{\Delta(t)}{2} \right )=R+\sqrt{1-R^2}\tan\left ( \frac{\left \langle\Omega\right \rangle^{12}   \sqrt{1-R^2}}{2}t+\tan^{-1}\left ( \frac{\tan\left ( \frac{\Delta(0)}{2} \right )-R}{\sqrt{1-R^2}} \right )\right ).
\label{eq:super_general_approx_no_conver}
\end{equation}

\noindent As can be observed, Eq.~\ref{eq:super_general_approx_no_conver} shows that $\tan\left ( \frac{\Delta(t)}{2} \right )$ is a periodic function for $\Lambda_{\Delta} \leq \left |\left \langle\Omega\right \rangle^{12} \right |$. This drifting behavior just states that, if the interlayer coupling strength is not large enough, it is no longer possible to reduce the difference of average frequencies between the layers and entrain the whole system.

Supposing that $\Delta/2 \gtrsim 0$, $\tan\left ( \frac{\Delta(t)}{2} \right ) \geq 2\left | R \right |$ and $\Lambda_{\Delta}\gg \left |\left \langle\Omega\right \rangle^{12}  \right |$, the absolute value signs in Eq.~\ref{eq:super_general_approx} can be removed and, thus, it can be approximated as:

\begin{equation}
\frac{\tan\left ( \frac{\Delta}{2} \right ) }{\tan\left ( \frac{\Delta}{2} \right ) - A}=-\frac{1}{A}\;\left ( \frac{\Delta}{2} \right )-\frac{1}{A^2}\;\left ( \frac{\Delta}{2} \right )^2-\frac{(A^2+3)}{3A^3}\;\left ( \frac{\Delta}{2} \right )^3-\dots \approx \xi(0) e^{-\Lambda_{\Delta} t},
\label{eq:taylor_xi}
\end{equation}

\noindent where $A=2\left | R \right |\mathrm{sgn}\left (\left \langle\Omega\right \rangle^{12}\right ).$ Under these conditions, the relaxation time of $\Delta$ is dominated once again by $\Lambda_{\Delta}$. Hence, provided that $r_1(t)\approx r_2(t) \approx 1$ and $\Lambda_{\Delta}\gg \left |\left \langle\Omega\right \rangle^{12}  \right |$, the non-linear Kuramoto model (Eq.~\ref{eq:kuramoto}) and the spectral analysis (see subsection \ref{Sec:relax_layers}) lead to the same relaxation time for the interlayer synchronization process for $M=2$: $\tau_{\Delta} =1/\Lambda_{\Delta}=1/2\lambda^{12}$.


For small values of $\Delta$, the time evolution of the order parameter in Eq.~\ref{eq:global_order_timescale} can be approximated by $r(t)\simeq 1- \Delta ^2/8$. Therefore, the timescale of the order parameter ($\tau_r$) is determined by the smallest nonzero power of $\Delta/2$, and a rough estimation is $\tau_r \gtrsim 1/2\Lambda_{\Delta}$.

Summarizing the results in sections \ref{Sec:theory} and \ref{Sec:relax_layers}, the asymptotic synchronization phase of the Kuramoto model on multiplexes is characterized by the following behavior:
\begin{enumerate}[label=(\roman*)]
\item When $\lambda^{12} \ll \lambda^{1}=\lambda^{2}$, the timescales
rank as follows: $\tau_{\mathcal{M}} = \tau_{\Delta} > \tau_{r}$.
\item When $\lambda^{12} \gg \lambda^{1}=\lambda^{2}$, the timescales
rank as follows: $\tau_{\mathcal{M}} \gg \tau_{\Delta} > \tau_{r}$.
\end{enumerate}

According to Eq.~\ref{eq:general_approx}, increasing the value of $\lambda^{12}$ accelerates the transient regimes of the interlayer synchonization and of the global order parameter, respectively. Additionally, it reduces the difference between the average phase of each layer and, hence, it favors the full synchronization of the system. The important aspect of this result is that, contrary to what is observed for the multiplex diffusive relaxation, when  $r_\alpha \simeq 1$.

These results are in accordance with the prior findings on \textit{superdiffusion} \cite{gomez13,sole13,serrano17}. Superdiffusion emerges when the timescale of the multiplex is faster than that of both layers acting separately \cite{gomez13, sole13}, i.e. $\Lambda_2 > \max(\sigma_{2}^{1},\sigma_{2}^{2})$, where $\sigma_{2}^{\alpha}$ is the smallest nonzero eigenvalue of the Laplacian matrix of layer $G_{\alpha}$. For large coupling between layers, spectral analysis predicts that superdiffusion is not guaranteed; it depends on the specific structures coupled together. Increasing the interlayer coupling accelerates the convergence of the global order parameter and of the difference between the average phase of each layer. Nevertheless, it also increases the magnitude of the pertubations that are transmitted across the interlayer.


\section{Numerical results}
\label{Numerical_results}
In this section we show that the prior analytical findings are in complete agreement with computer simulations. We compare results of the numerical integration of the coupled Kuramoto oscillators for several multiplexes realizations, using 16 digit variables. From the solution for $\theta_n^\alpha(t)$ we obtain the time evolution of $\tan\left ( \frac{\Delta(t) }{2} \right )$ and $1-r\left ( t \right )$ for the linear and non-linear regimes that are compared, respectively, to

\begin{eqnarray}
\nonumber  \eta_2(t) &=& \left | \tan\left ( \frac{\Delta(0) }{2} \right ) \right |e^{-\Lambda_2t}, \\
  \eta_r(t) &=& \left ( 1-r\left ( 0 \right ) \right )e^{-2\Lambda_{\Delta}t}. 
  \label{eq:decay_lambda_delta}
\end{eqnarray}

\noindent $\eta_r(t)$ is a measure of the synchronization dynamics, while 
$\eta_2(t)$ has the same dependence on time as the multiplex diffusive dynamics. Besides that, $\tan\left ( \frac{\Delta(t) }{2} \right )$ is also compared to $\eta_\Delta(t)$ in Eq.~\ref{eq:general_approx}.

Other examples for different values of the interlayer and intralayer coupling constants and several initial conditions for the coupled Kuramoto oscillators, are presented in the Supplementary Material to this paper. All of them are in complete agreement with the results described in this section.

\subsection{Linear Kuramoto model}
\label{linear_results}

We start by presenting numerical results from the integration of Eq.~\ref{eq:diffusion}, where the initial phases $\theta_n^{\alpha}(0)$ are drawn randomly from a uniform distribution $\mathcal{U_{\theta^{\alpha}}}\left ( \mu_{\alpha}-a,\mu_{\alpha}+a\right )$, and $\mu_{\alpha}$ is the expected value of $\theta_n^{\alpha}$. Results satisfying $a\ll1$ can be compared to the analytical expressions derived in the previous sections for $\tan\left ( \frac{\Delta}{2} \right )$ and $1-r$, as in these cases the condition $r_\alpha\simeq 1$ is satisfied. For the sake of an easier comparison with the analytical results, we set  $\lambda^1=\lambda^2=\lambda$. We remark that results depend on the following factors: coupling strengths, initial conditions and network topology.

\begin{figure}[h!]
\centering
\includegraphics[width=1.0\textwidth]{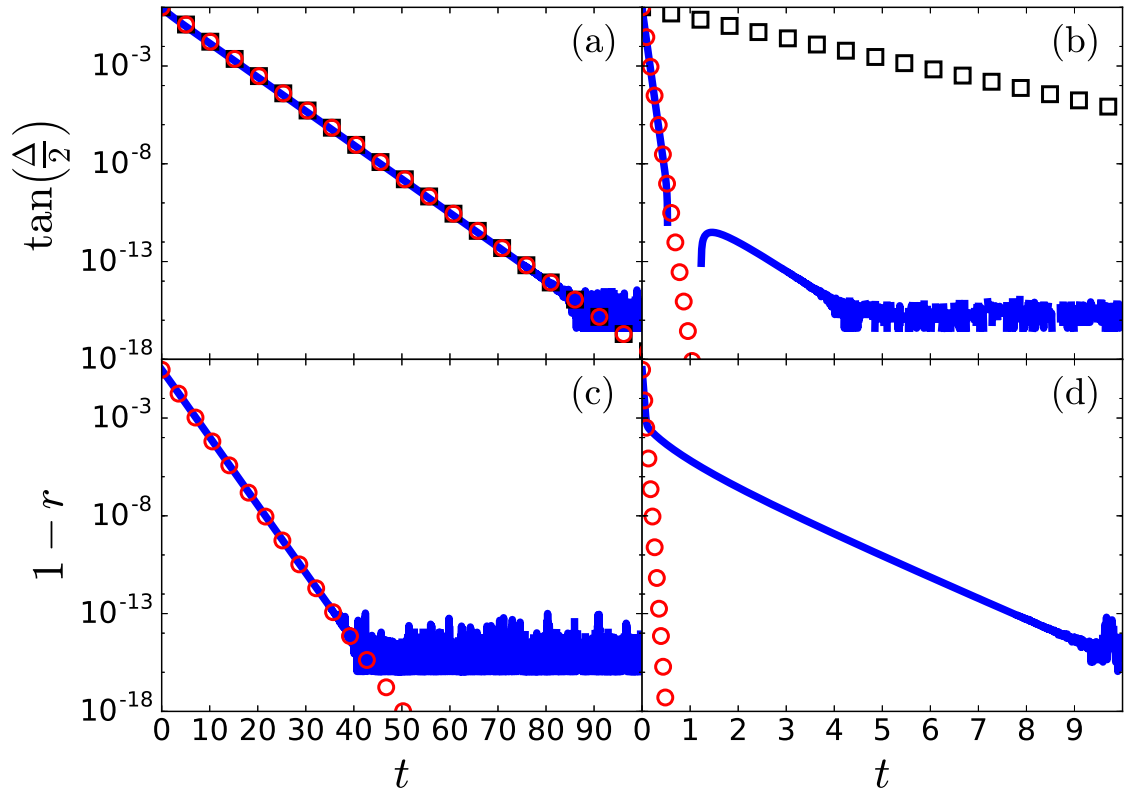}
\caption{Numerical results for $N=500$, $\lambda=2.0$, $\mu_{1}=\pi/2$, $\mu_{2}=0$, and $a=0.1$. Each multiplex layer has the same topological features described in Fig.~\ref{fig:evolucion_Lambda_23}. Panels (a) and (b): Time evolution of $\tan\left ( \frac{\Delta(t) }{2} \right )$ (blue continuous line), $\eta_{\Delta}(t)$ (red circles) and $\eta_2(t)$ (black squares) for $\lambda^{12}=0.1\lambda$ (a), and $\lambda^{12}=10.0\lambda$ (b). Panels (c) and (d): Time evolution of $1-r\left ( t \right )$ (blue continuous line) and $\eta_r(t)$ (red circles) for $\lambda^{12}=0.1\lambda$ (c), and $\lambda^{12}=10.0\lambda$ (d).}
\label{new_3}
\end{figure}





Dependence on coupling strengths is in agreement with section \ref{Sec:relax_layers}. Fig.~\ref{new_3}a shows that, for $\lambda^{12}\ll\lambda$, the timescales of interlayer synchronization and of diffusion on $\mathcal{M}$ are equal: the time evolution of $\tan\left ( \frac{\Delta(t) }{2} \right )$ is well approximated by $\eta_{\Delta}(t)$  and $\eta_2(t)$ , i.e. $\Lambda_{2} \approx \Lambda_{\Delta}$. However, when $\lambda^{12}\gg\lambda$, these timescales differ, i.e. $\Lambda_{2} \neq \Lambda_{\Delta}$, as indicated by lines with different slopes in Fig.~\ref{new_3}b. Moreover, it is also shown that the agreement between $\tan\left ( \frac{\Delta(t) }{2} \right )$ and  $\eta_{\Delta}(t)$ has a lower limit $\sim 10^{-10}$. Nevertheless, the difference between the average phases of both layers relaxes faster than the whole system,  i.e. $\tau_{\mathcal{M}} \gg \tau_{\Delta}$ for $\lambda \ll \lambda^{12}$. Both panels reveal the presence of random fluctuations $\sim 10^{-15}$, which depend on precision of the used variables.

The same (somewhat different) features are observed in Figs.~\ref{new_3}c (Fig.~\ref{new_3}d), where we compare the approximation $\eta_r(t)$ with the actual value of $1-r\left ( t \right )$. The evolution of  $1-r\left ( t \right )$ is well adjusted by $\eta_r(t)$ for $\lambda^{12}\ll \lambda$. However, when $\lambda^{12}\gg \lambda$, the quantities agree with each other in a more limited range $\gtrsim 10^{-4}$ .


For a given choice of the coupling parameters, the deviations from the exponential behavior can be influenced by topological differences among the layers and by the initial values $\theta_n^{\alpha}(0)$. To emphasize the importance of the later, we consider $M=2$ multiplexes where each layer consists of a complete graph, for which make analytical expressions for $\Lambda_2$ can be obtained (see Appendix). In Fig.~\ref{fig_b:LINEAL__CON_THETA_0.1} we show the numerical results for $1-r(t)$ when $a=0$ and $0.1$. The inset shows that the time evolution of $1-r\left ( t \right )$ is well adjusted by $\eta_r(t)$, when $a=0$, while departures from the exponential decay take place when $a>0$. Here, the agreement between the curves is limited to the range $\gtrsim10^{-6}$.

\begin{figure}[!htb]
    \centering
        {\includegraphics[width=0.6\textwidth]{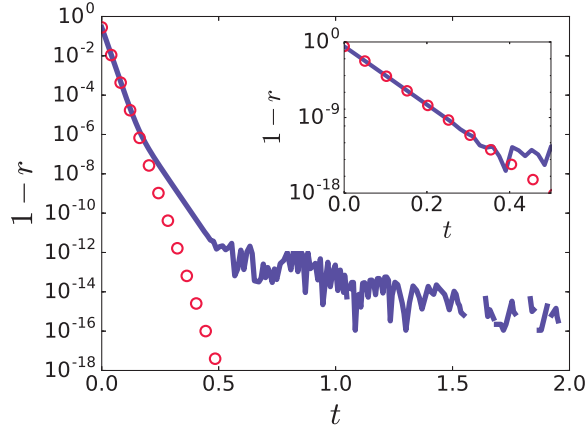}}
    \caption{Time evolution of $1-r\left ( t \right )$ (blue continous line) and $\eta_r(t)$ (red circles) for $N=10$, $\lambda=2.0$, $\lambda^{12}=10\lambda$, $\mu_{1}=\pi/2$, $\mu_{2}=0$ and $a=0.1$. Each layer contains a complete graph. The inset shows the results by considering $a=0$.}
\label{fig_b:LINEAL__CON_THETA_0.1}
\end{figure}


Fig.~\ref{new_3} and Fig.~\ref{fig_b:LINEAL__CON_THETA_0.1} suggest that it may be possible to relate the range of values of $1-r$ where the numerical results coincide with the analytical predictions to $\tau_D$, the characteristic timescale for the emergence of these discrepancies. It turns out that $\tau_D$  is mainly controlled by the value of $\Lambda_2$ as follows:

\begin{equation}
\tau_D\approx \frac{1}{2\Lambda_2}.
\label{discrep_time}
\end{equation}

\noindent Therefore, in case $\Lambda_{\Delta} \approx \Lambda_2$, deviations disappear until the numeric precision of the used variables is reached, whether or not $a=0$  (see Fig.~\ref{new_3}a and Fig.~\ref{new_3}c). However,  if $\Lambda_{\Delta}>\Lambda_{2}$ and $a>0$, discrepancies will manifest.

\begin{figure}[h!]
\centering
\subfloat[]{\label{fig_b:LINEAL_ERROR_10_NODES_EFECT}\includegraphics[width=0.5\textwidth]{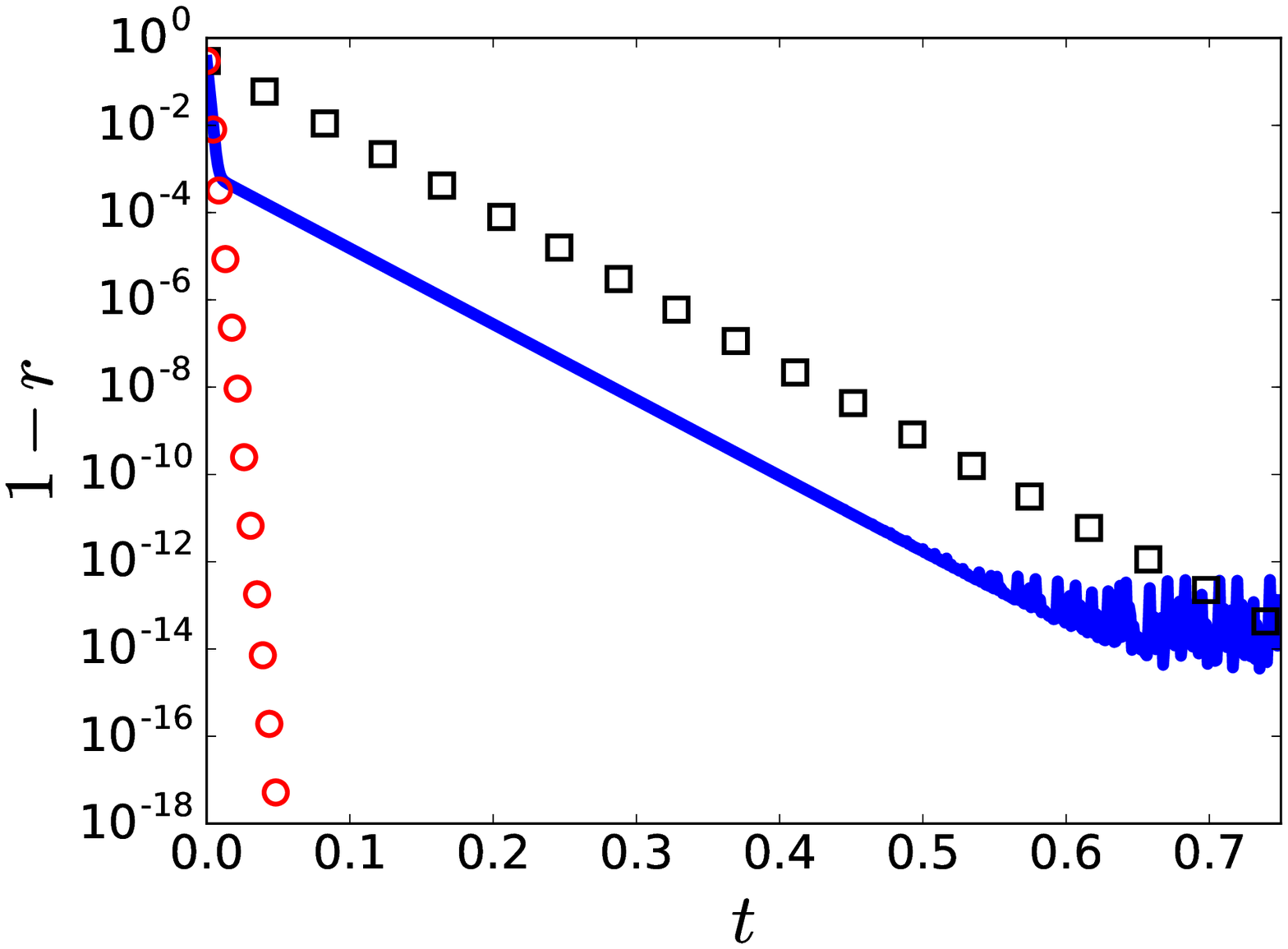}}
\subfloat[]{\label{fig_b:LINEAL_ERROR_100_NODES_EFECT}\includegraphics[width=0.5\textwidth]{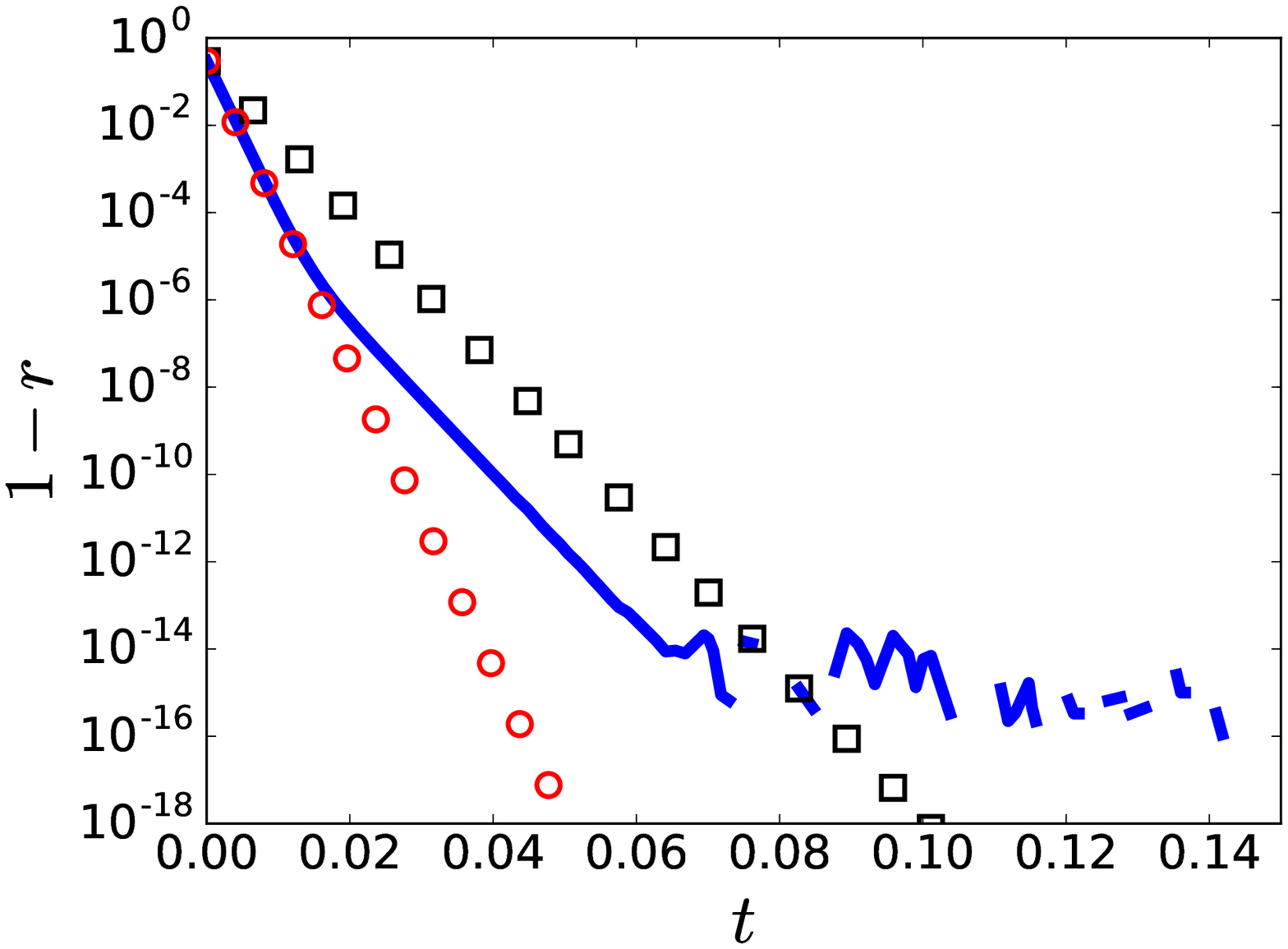}}
\caption{Time evolution of $1-r\left ( t \right )$ (blue continuous line), $\eta_r(t)$ (red circles) and a guide for the eye proportional to $e^{-2\lambda Nt}$ (black squares), for $\lambda=2.0$, $\lambda^{12}=100\lambda$, $\mu_{1}=\pi/2$, $\mu_{2}=0$ and $a=0.1$. Each layer contains a complete graph. (a) Left panel:  $N=10$ (b) Right panel: $N=100$.}
\end{figure}

Finally, still using complete graphs for the sake of comparison to analytical expressions, we illustrate the dependence of the multiplex dynamics on the topology, for a given choice of the coupling strengths and the initial conditions. We note that the dependence on the topology can be observed just by changing the number of nodes in each layer of complete graph. Indeed, if $\Lambda_{\Delta}>\Lambda_{2}$, the smallest nonzero eigenvalue of the supra-Laplacian matrix is $\Lambda_2=\lambda N$ (see Appendix). Therefore, according to Eq.~\ref{discrep_time}, the smaller the number of nodes $N$, the larger the desviations, for $\tau_{\mathcal{M}}>\tau_{\Delta}$ and $a>0$. In Fig.~\ref{fig_b:LINEAL_ERROR_10_NODES_EFECT} and Fig.~\ref{fig_b:LINEAL_ERROR_100_NODES_EFECT} we display the time evolution of $1-r\left ( t \right )$, $\eta_r(t)$, and a guide for the eye proportional to $e^{-2\lambda Nt}$ for $N=10$ and $N=100$, respectively, and $a>0$. As can be observed, these results are in good agreement with Eq.~\ref{discrep_time}. In the Appendix, we show analytically the dependence of the global order parameter $r$ on $e^{-2\Lambda_2t}$ (i.e. $e^{-2\lambda Nt}$), when each layer of the multiplex network is a complete graph.

\subsection{Non-linear Kuramoto model}
\label{results_Non-linear}

\begin{figure}[h!]
\centering
\subfloat[]{\label{NOLINEAL_ANG_FUERTE_SIN_FREC}\includegraphics[width=0.5\textwidth]{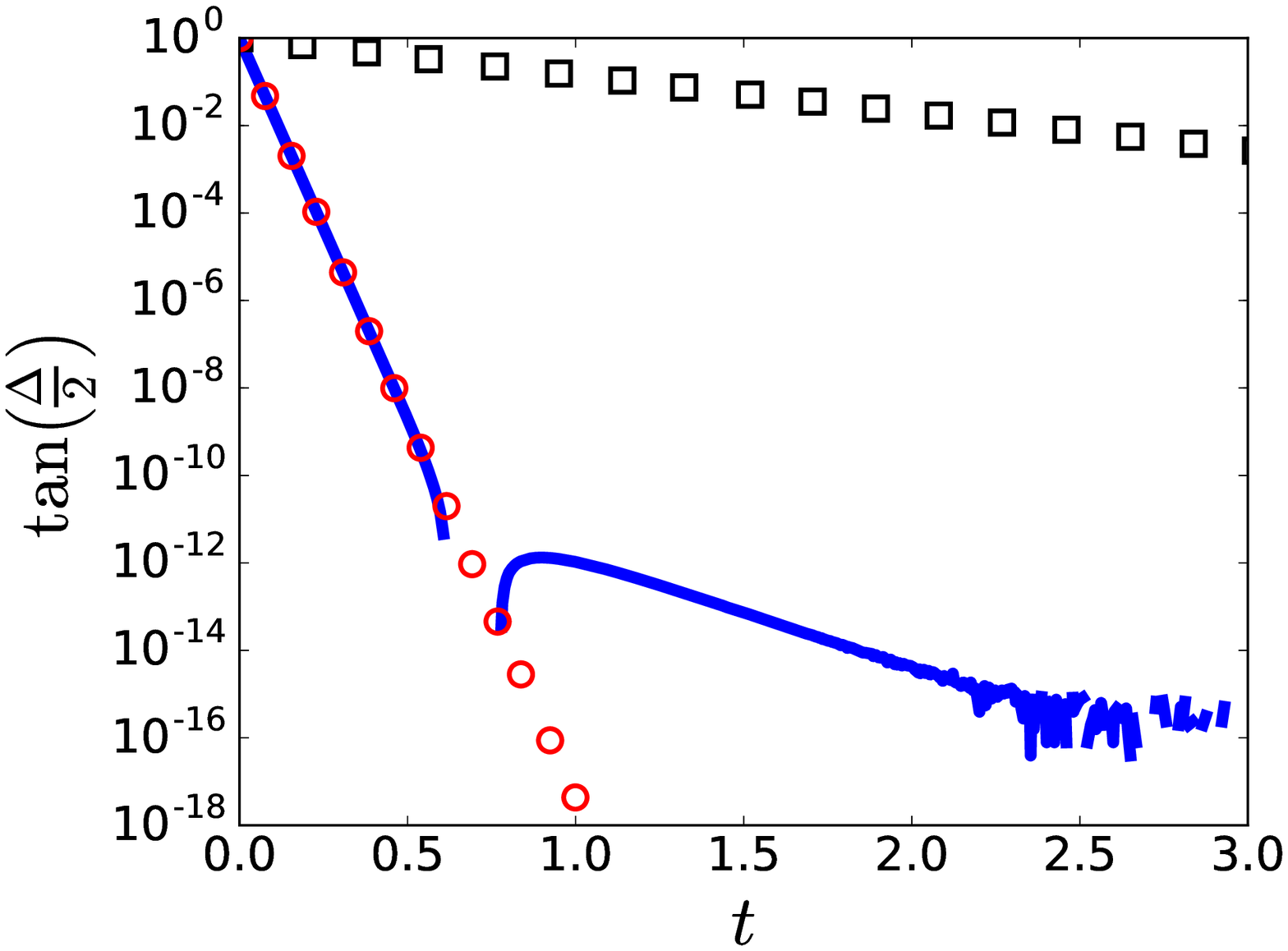}}
\subfloat[]{\label{NOLINEAL_ERROR_FUERTE_SIN_FREC}\includegraphics[width=0.5\textwidth]{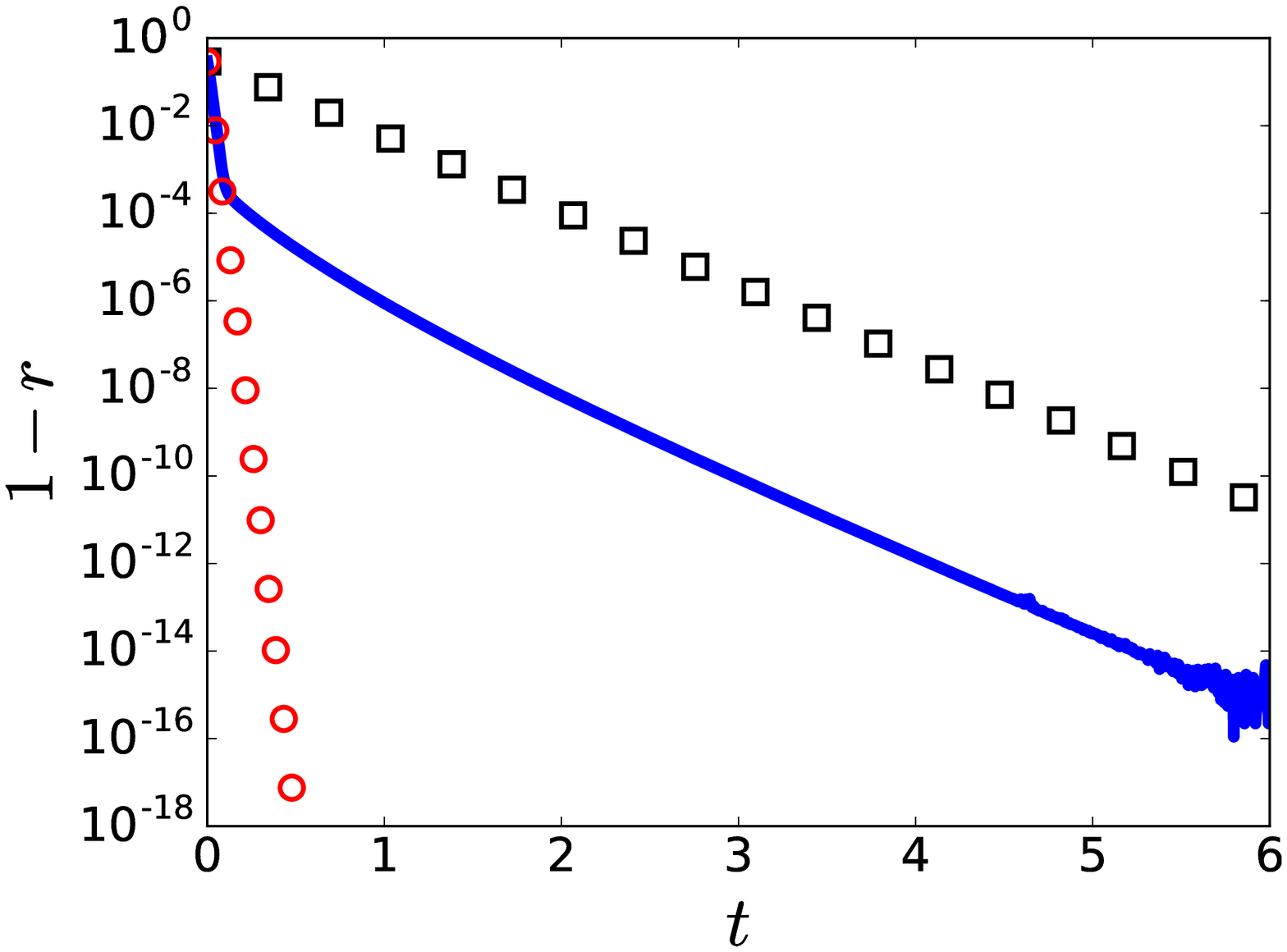}}
\caption{(a) Left panel: Time evolution of $ \tan\left ( \frac{\Delta(t) }{2} \right )$, $\eta_{\Delta}(t)$, and $\eta_2(t)$. (b) Right panel: Time evolution of $1-r\left ( t \right )$ and $\eta_r(t)$. $\lambda^{12}=10.0\lambda$ in both panels, and the used symbols and lines are the same as in Fig.~\ref{new_3}b and Fig.~\ref{new_3}d. The multiplexes are the same as those used in Fig.~\ref{new_3}.  }
\end{figure}

The numerical results for the non-linear equations Eq.~\ref{eq:kuramoto} were obtained using the same procedure described in previous subsection. When all natural frequencies of the oscillators are set to zero, i.e. $\Omega_n^{\alpha}=0$ $\forall n$, the time evolution of $\tan\left ( \frac{\Delta(t) }{2} \right )$ and $1-r(t)$ for $\lambda ^{12} \ll \lambda$ are essentially the same as those in Fig.~\ref{new_3}a and Fig.~\ref{new_3}c.
However, when $\lambda ^{12} \gg \lambda$, which causes  $\Lambda_{2} \neq \Lambda_{\Delta}$ and $\tau_{\mathcal{M}} \gg \tau_{\Delta}$, $\tan\left ( \frac{\Delta(t) }{2} \right )$ deviates both from  $\eta_2(t)$ and $\eta_\Delta(t)$ as well as $1-r(t)$ deviates from $\eta_r(t)$. The comparison between Fig.~\ref{new_3}b and Fig.~\ref{NOLINEAL_ANG_FUERTE_SIN_FREC} shows that the non-linear terms affects the evolution $\tan\left ( \frac{\Delta(t) }{2} \right )$. Notice that the effect on the evolution of $1-r(t) \sim \Delta^2$ is much smaller, in such a way that the changes induced by the non-linear terms in Fig.~\ref{NOLINEAL_ERROR_FUERTE_SIN_FREC} are minute in comparison to Fig.~\ref{new_3}d.

Dependence of $1-r(t)$ on $a$ for $M=2$ multiplexes formed by complete graphs is very similar to that in Fig.~\ref{fig_b:LINEAL__CON_THETA_0.1}. When $a=0$, $1-r\left ( t \right )$ and  $\eta_r(t)$ are in complete agreement, if they are greater or simmilar to $10^{-12}$; while for $a=0.1$ deviations appear when $\eta_r(t)\lesssim 10^{-5}$.

\begin{figure}[h!]
\centering
\subfloat[]{\label{fig_b:multiplex_ac_MUY_debil}\includegraphics[width=0.5\textwidth]{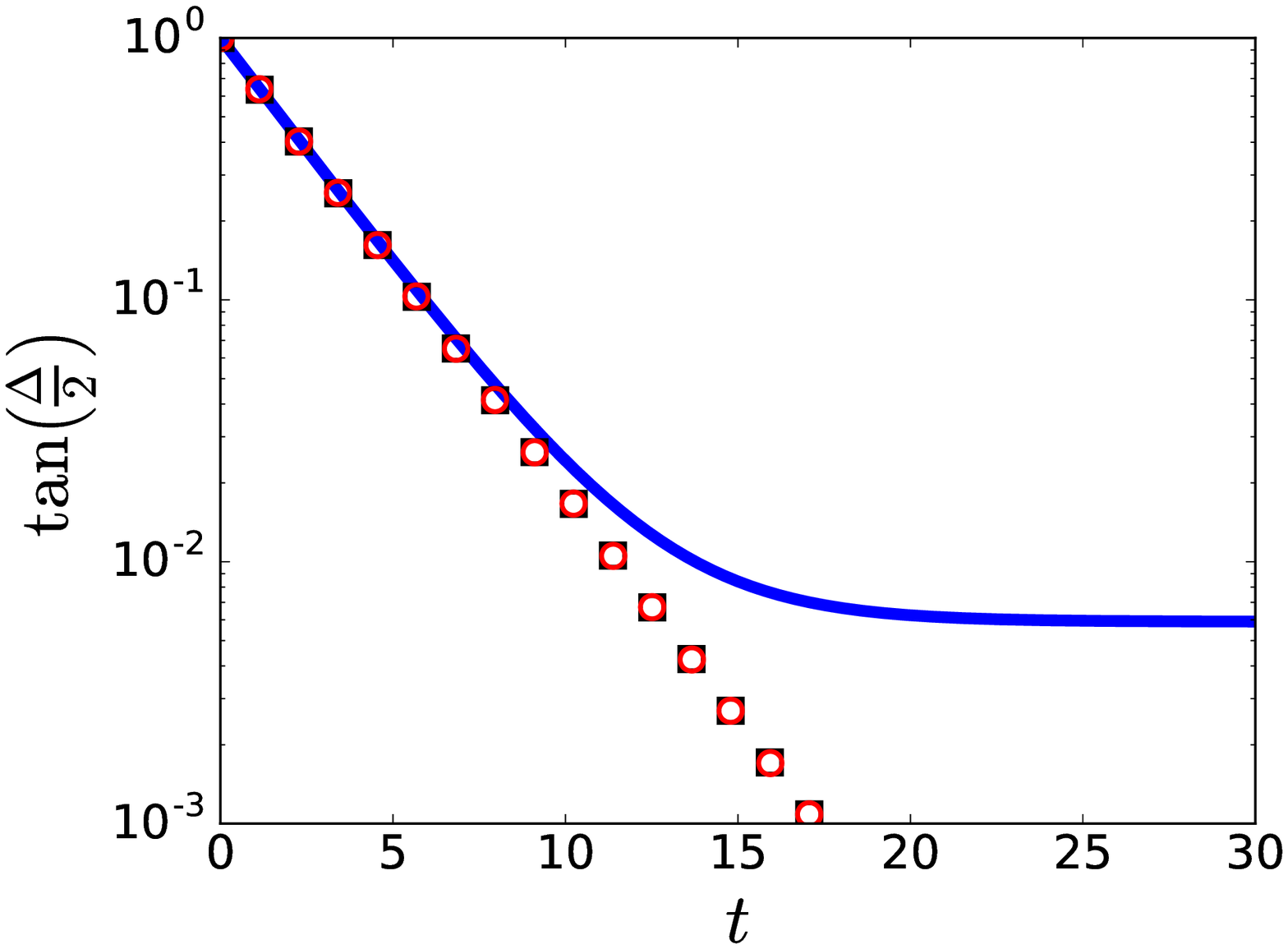}}
\subfloat[]{\label{fig_b:multiplex_ac_MUY_fuerte}\includegraphics[width=0.5\textwidth]{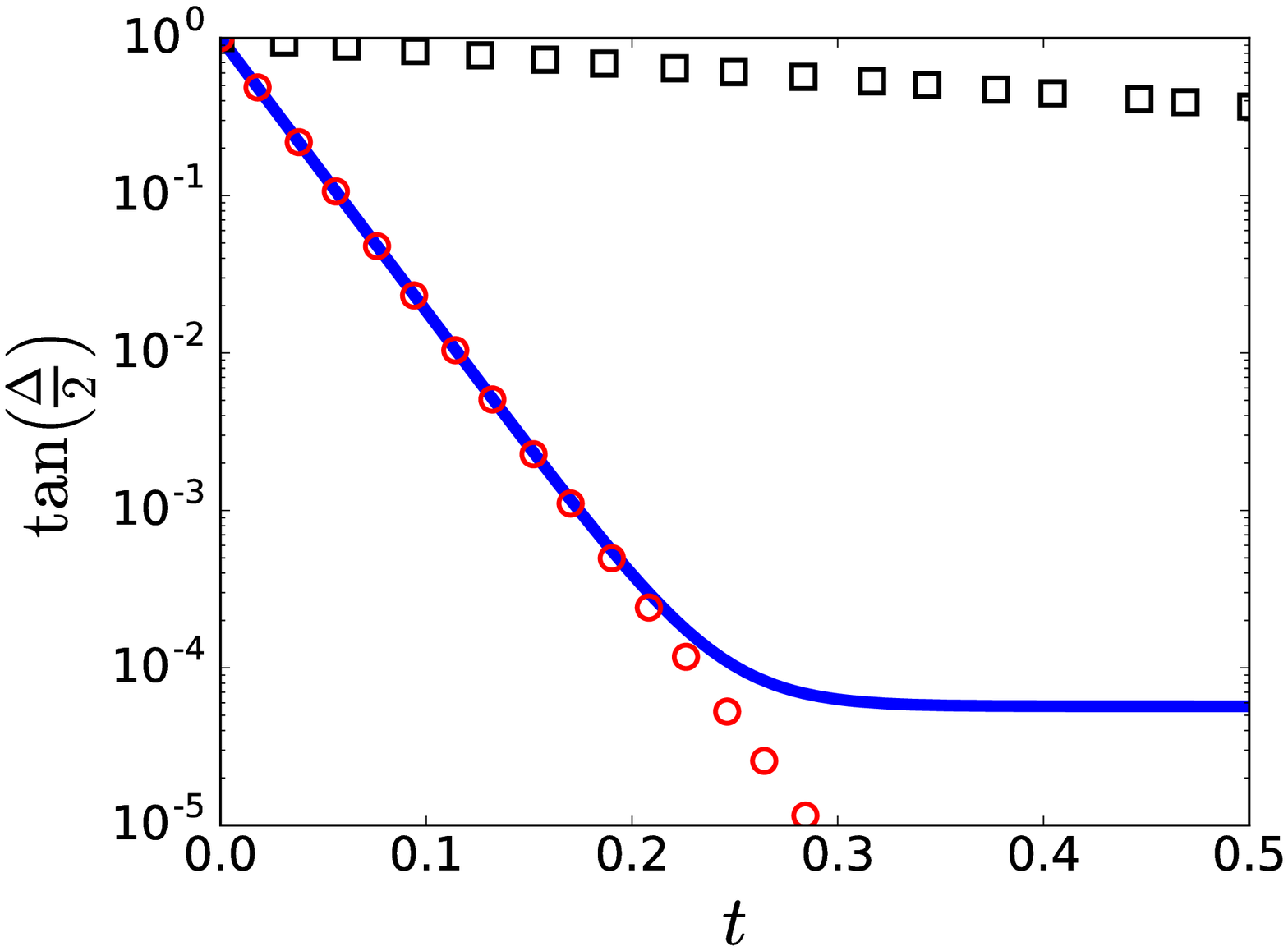}}
\caption{Time evolution of $ \tan\left ( \frac{\Delta(t) }{2} \right )$, $\eta_{\Delta}(t)$, and $\eta_2(t)$. The multiplex parameters, symbols and lines are the same as in Fig.~\ref{new_3}a and Fig.~\ref{new_3}b, except for $\Omega_n^{\alpha} \in \mathcal{U}\left ( 0.8,1.2\right )$. (a) Left panel: $\lambda^{12}=0.1\lambda$ (b) Right panel: $\lambda^{12}=10.0\lambda$.}
\end{figure}

Let us now discuss the results when the natural frequencies $\Omega_n^{\alpha}$ are different from zero so that, in general, $\left \langle\Omega\right \rangle_{1}\neq\left \langle\Omega\right \rangle_{2}$. Following \cite{avalos12}, the values of the frequencies are drawn randomly from a uniform distribution $\mathcal{U}\left ( 0.8,1.2\right )$. As observed in Fig.~\ref{fig_b:multiplex_ac_MUY_debil} and Fig.~\ref{fig_b:multiplex_ac_MUY_fuerte}, the time evolution of $\tan\left ( \frac{\Delta(t) }{2} \right )$ diverges from $\eta_{\Delta}(t)$ when $\left \langle\Omega\right \rangle_{1}\neq\left \langle\Omega\right \rangle_{2}$, for both $\lambda ^{12} \ll \lambda$ and $\lambda^{12} \gg \lambda$. In both cases $\Delta$ converges to a non-zero value and, consequently, the oscillators do not reach a full synchronization in accordance to Eq.~\ref{eq:super_general_approx} and Eq.~\ref{eq_R}. We notice that the deviations from the exponential predictions for $\lambda ^{12} \ll \lambda$ occur at a larger value of $\eta_2(t)$ as compared to $\lambda ^{12} \ll \lambda$. This stays in opposition to the previously observed behavior for $\Omega_n^{\alpha}\equiv0$. Indeed, a relatively small interlayer coupling favors the emergence of the deviations, once interlayer synchronization is impeded for $\lambda^{12}\approx 0$. Hence, if $\left |\left \langle\Omega\right \rangle^{12}  \right |>0$ and $\lambda^{12}\approx 0$, the exponential decay barely takes place. In the case of $\lambda^{12}\gg 0$, the relaxation time of the synchronization error gets closer to the estimation given by $\eta_r(t)$, whether or not $\lambda \gg \lambda^{12}$.

The asymptotic value of the difference between the average phases of both layers can be
estimated from Eq.~\ref{eq:super_general_approx}. If $\tan\left ( \frac{\Delta(t)}{2} \right )\geq \mathrm{sgn}\left (\left \langle\Omega\right \rangle^{12} \right )\left ( \left | R \right |+\sqrt{R^2-1} \right ) $,  Eq.~\ref{eq:super_general_approx} can be rewritten as 

\begin{equation}
\tan\left ( \frac{\Delta(t)}{2} \right )=
\left ( \left | R \right | -\sqrt{R^2-1}\;\frac{1+\xi(0)e^{-t\left |\left \langle\Omega\right \rangle^{12}  \right |\sqrt{R^2-1}}}{1-\xi(0)e^{-t\left |\left \langle\Omega\right \rangle^{12}  \right |\sqrt{R^2-1}}} \right )\mathrm{sgn}\left (\left \langle\Omega\right \rangle^{12}  \right),
\end{equation}

\noindent so that its asymptotic value $t\rightarrow\infty$ is given by

\begin{equation}
\lim_{t\rightarrow \infty }
\tan\left ( \frac{\Delta(t)}{2} \right )=
\left ( \left | R \right | -\sqrt{R^2-1} \right )\mathrm{sgn}\left (\left \langle\Omega\right \rangle^{12} \right ).
\label{dif_asintotica}
\end{equation}

\begin{figure}[h!]
\centering
\includegraphics[width=0.5\textwidth]{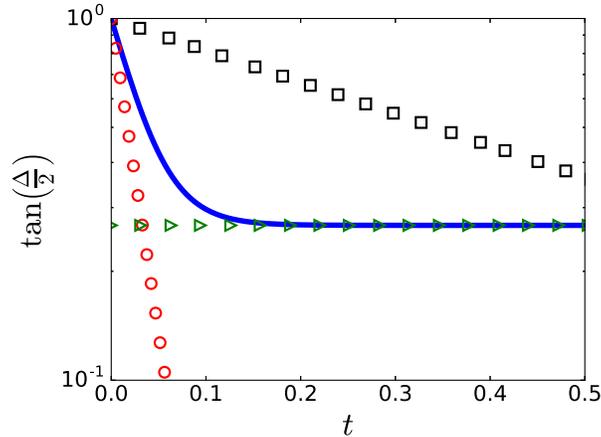}
\caption{Time evolution of $ \tan\left ( \frac{\Delta(t) }{2} \right )$, $\eta_{\Delta}(t)$, and $\eta_2(t)$. The multiplex parameters, symbols and lines are the same as in Fig.~\ref{new_3}b. 
The model parameters are $\lambda=2.0$, $\lambda^{12}=10\lambda$ and $2\left \langle\Omega\right \rangle^{12}=\Lambda_{\Delta}$.  Green triangles indicate the asymptotic value obtained with Eq.~\ref{eq:super_general_approx}.}
\label{fig_b:multiplex_CON_FRECUENCIAS}
\end{figure}

If $\left \langle\Omega\right \rangle_{1}\simeq \left \langle\Omega\right \rangle_{2}$, $R$ diverges and $\Delta$ decays to zero exponentially. On the other hand, in Fig.~\ref{fig_b:multiplex_CON_FRECUENCIAS} we expose the time evolution of $\tan\left ( \frac{\Delta(t) }{2} \right )$ for $2 \left \langle\Omega\right \rangle^{12}=\Lambda_{\Delta}$. In that case, according to Eq.~\ref{eq:super_general_approx} and  Eq.~\ref{dif_asintotica}, the asymptotic value of the difference between the average phases of both layers is $\psi^1-\psi^2=\pi/6$ (green triangles). It is easy to see that the prior estimation is very accurate.

\begin{figure}[h!]
\centering
\subfloat[]{\label{fig_b:NOLINEAL_ERROR_DEBIL}\includegraphics[width=0.5\textwidth]{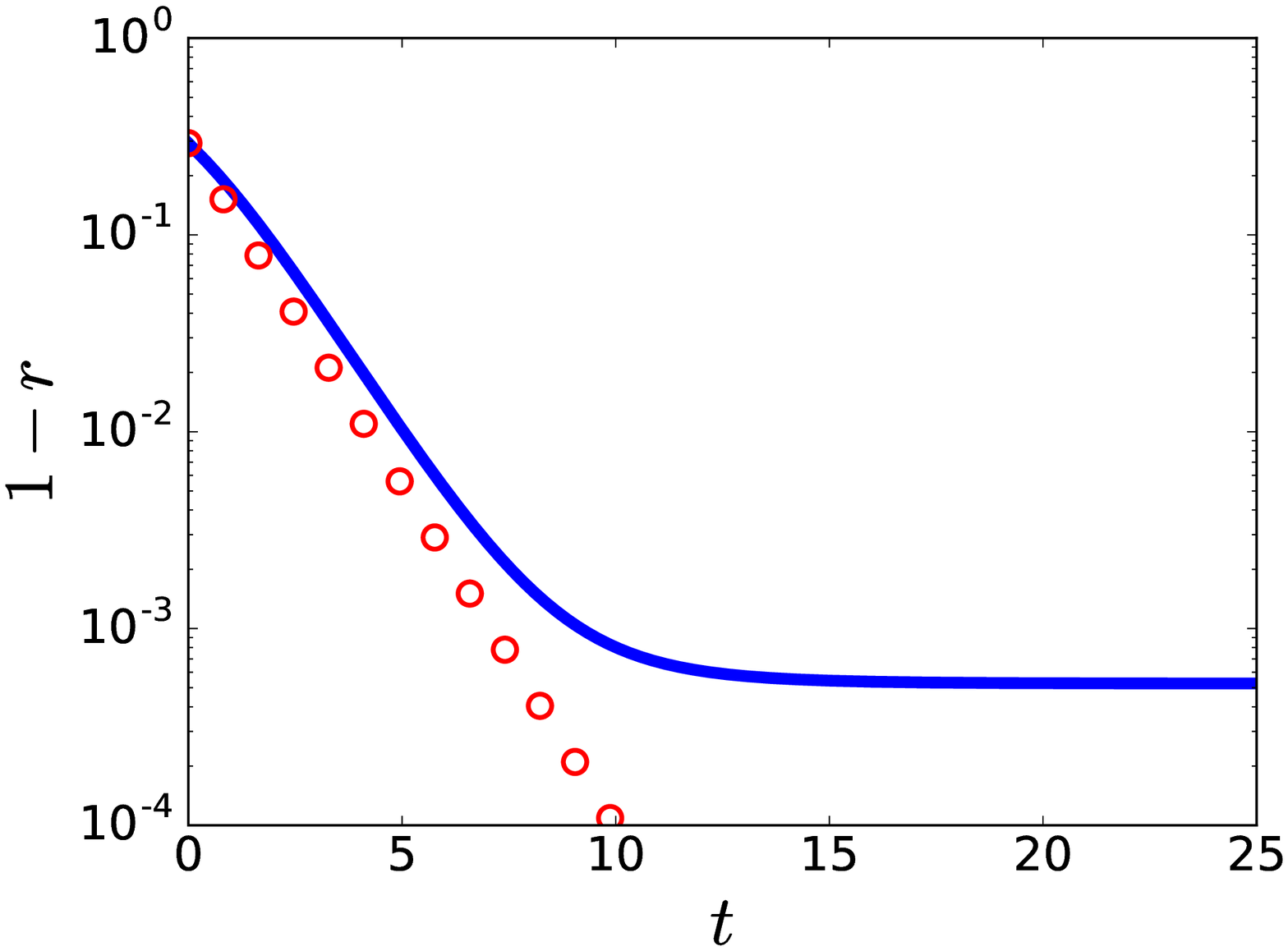}}
\subfloat[]{\label{fig_b:NOLINEAL_ERROR_FUERTE}\includegraphics[width=0.5\textwidth]{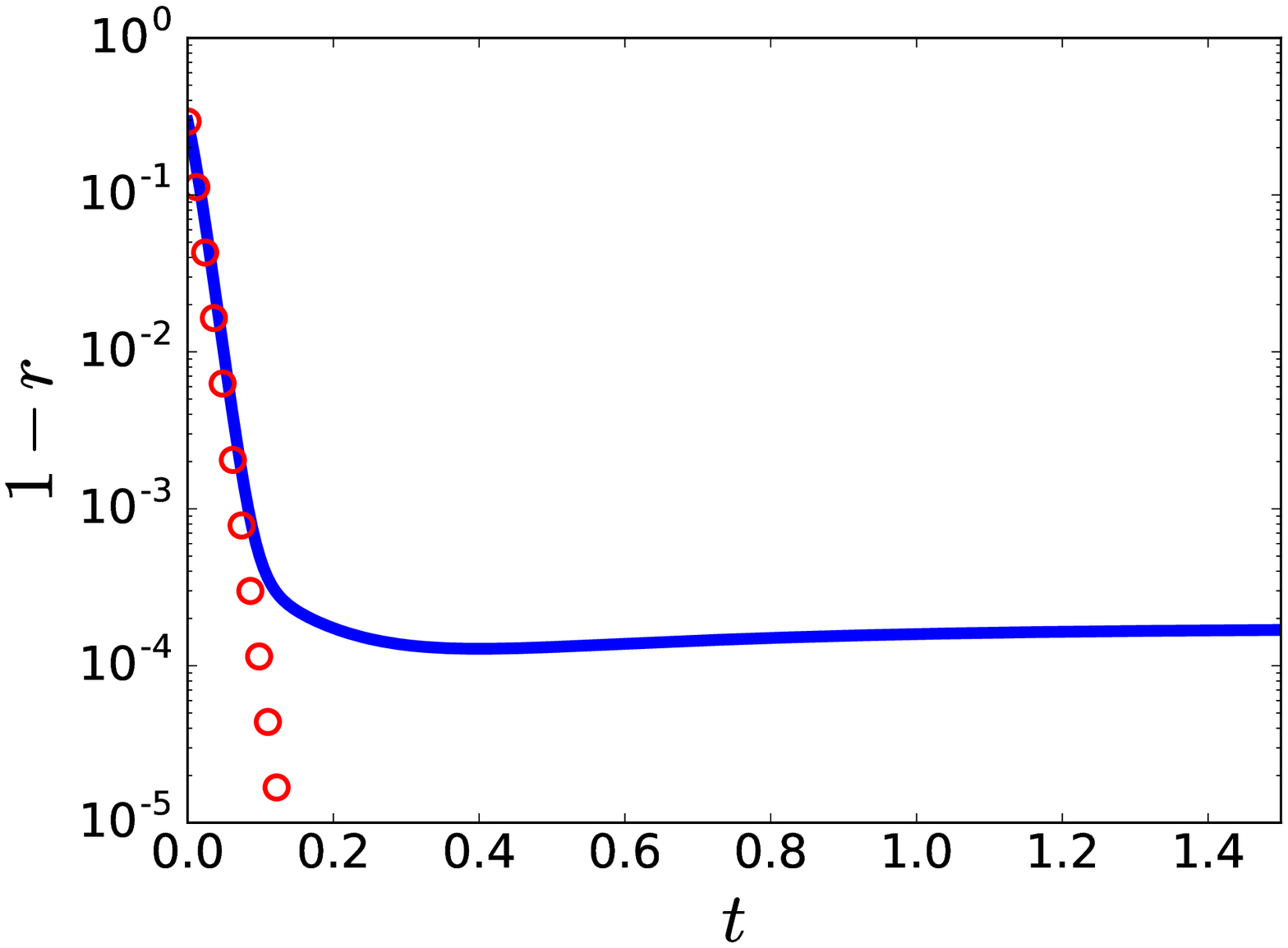}}
\caption{Time evolution of $1-r\left ( t \right )$ and $\eta_r(t)$. The multiplex parameters, symbols and lines are the same as in Fig.~\ref{new_3}c and Fig.~\ref{new_3}d, except for $\Omega_n^{\alpha} \in \mathcal{U}\left ( 0.8,1.2\right )$. (a) Left panel: $\lambda^{12}=0.1\lambda$ (b) Right panel: $\lambda^{12}=10.0\lambda$.}
\end{figure}

Fig.~\ref{fig_b:NOLINEAL_ERROR_DEBIL} and Fig.~\ref{fig_b:NOLINEAL_ERROR_FUERTE} illustrate the behavior of $1-r\left ( t \right )$ for small and large interlayer coupling, respectively. As can be observed, synchronization error departs from $\eta_r(t)$ values whether or not $\lambda ^{12} \ll \lambda$. As expected, its asymptotic value does not decays to zero.

\section{Conclusions}
\label{Conclusions}

We have developed a simple formalism to study
the timescales of the global order parameter and the interlayer synchronization of multilayer networks. Our approach has been addapted to a two-layer multiplex with high degrees of synchronization in each layer (i.e. $r_{\alpha}(t)\approx 1$ for $1 \leq \alpha \leq 2$ and $t\geq 0$), in a particular setup in which nodes are preserved through layers.

We have analyzed the difference between the average phase of each layer of the multiplex network from two different perspectives: spectral analysis and non-linear Kuramoto model. Our analytical results showed that the timescales of the global order parameter $\tau_r$ and the interlayer synchronization $\tau_{\Delta}$ are inversely proportional to the interlayer coupling strength $\lambda^{12}$. Surprisingly, the convergence of the global order parameter is faster than the convergence of interlayer synchronization, and the latter is generally faster than the relaxation time of the multiplex network $\tau_{\mathcal{M}}$. These features do not depend on the specific structures coupled together. Therefore, increasing the interlayer coupling always shortens the global order parameter and the interlayer synchronization transient regimes.

On the other hand, our formalism outlined the effects of frequencies on evolution of the global order parameter and on interlayer synchronization process. In addition, conditions for an oscillatory behavior were also identified.

The analytical findings were in fairly good agreement with computer simulations. In the case of multiplex networks with relatively small interlayer coupling (i.e. $\lambda^{12}\ll \lambda$), similar average frequencies in each layer (i.e. $\left \langle\Omega\right \rangle_{1}\approx\left \langle\Omega\right \rangle_{2}$) and high degrees of synchronization in each layer, at the initial time (i.e. $r_{\alpha}(0)\approx 1$ for $1 \leq \alpha \leq 2$),  analytical results and numerical ones were in complete agreement. However, supposing similar average frequencies in each layer, if the interlayer coupling is relatively large (i.e. $\lambda^{12}\gg \lambda$), and there exists an initial intralayer phase heterogeneity (i.e. there is at least one layer $G_{\alpha}$ that contains two or more oscillators whose phases are different at $t=0$), numerical results showed deviations from the predicted exponential decay, although major changes of the global order parameter and of the interlayer synchronization were fairly well adjusted by our analytical approach. The timescale of these discrepancies $\tau_D$ is inversely proportional to twice the smallest non-zero eigenvalue of the supra-Laplacian matrix $\mathcal{L}$ of the multiplex network, $\Lambda_2$. According to prior works \cite{gomez13,sole13}, this dependence on $\Lambda_2$ implies that deviations from our analytical results are shaped by topological characteristics of the layers involved as well as the respective values of $\lambda$ and $\lambda^{12}$.

When the average frequencies of each layer are dissimilar (i.e. $\left \langle\Omega\right \rangle^{12}=\left \langle\Omega\right \rangle_{1}-\left \langle\Omega\right \rangle_{2}\neq0$), computer simulations are in good agreement with our analytical results. If $\Lambda_{\Delta} \geq \left |\left \langle\Omega\right \rangle^{12} \right |$, the asymptotic values of the global order parameter and of the interlayer synchronization converge to a non-zero value. If $\Lambda_{\Delta} \leq \left |\left \langle\Omega\right \rangle^{12} \right |$, a periodic behaviour is obtained. Discrepancies from our analytical description do not appear, unless the asymptotic values of the global order parameter and of the interlayer synchronization are close to zero (i.e. $\left \langle\Omega\right \rangle^{12}\approx 0$).

Thus, under the hypotheses of this work, we conclude that timescale of the global order parameter is at least half times smaller than timescale of multiplex networks (i.e. $2\tau_r\approx 2\tau_D \approx\tau_{\mathcal{M}}=1/\Lambda_2$) and the major changes of this parameter are fairly well adjusted by our analytical findings (i.e. $\tau_r\approx\tau_{\Delta}=1/\Lambda_{\Delta}=1/2\lambda^{12}$).


\begin{acknowledgments}

This work was supported by the project MTM2015-63914-P from the Ministry of Economy and Competitiveness of Spain and by the Brazilian agency CNPq (grant 305060/2015-5). RFSA also acknowledges the support of the National Institute of Science and Technology for Complex Systems (INCT-SC Brazil).

\end{acknowledgments}


\section*{APPENDIX: ANALYTICAL RESULTS FOR A MULTIPLEX NETWORK FORMED BY COMPLETE GRAPHS.}

\subsection{Eigenvalue spectrum of the supra-Laplacian matrix.}
\label{App_espectro}

Given an undirected multiplex network $\mathcal{M}$ with $M=2$ layers, if both layers contain a complete network, then the supra-Laplacian matrix $\mathcal{L}$ has the following eigenvalues $\Lambda$:

\begin{enumerate}[label=(\roman*)]
\item $\Lambda=0$. It is a nondegenerate eigenvalue.
\item $\Lambda=\lambda N$. It is a degenerate eigenvalue. It appears $N-1$ times.
\item $\Lambda=2\lambda^{12}$. It is a nondegenerate eigenvalue.
\item $\Lambda=2\lambda^{12}+\lambda N$. It is a degenerate eigenvalue. It appears $N-1$ times.
\end{enumerate}

\noindent Thus, in case of $\lambda^{12}/\lambda \geq N/2$ ($\lambda^{12}/\lambda < N/2$), the smallest nonzero eigenvalue of the supra-Laplacian matrix is $\Lambda=\lambda N$ ($\Lambda=2\lambda^{12}$).


\subsection{Estimation of the average time evolution of the linear Kuramoto model.}
\label{teoria_complete}

Given an undirected multiplex network $\mathcal{M}$ with $M=2$ layers, if both layers contain a complete network, then Eq.~\ref{eq:diffusion} results in

\begin{equation}
\dot{\theta}_n^{\alpha}(t)=\lambda^{\alpha}N\left \langle \theta ^{\alpha}  \right \rangle-\lambda^{\alpha}N\theta_n^{\alpha}
+
\lambda^{12} \left ( \theta_n^{\beta}-\theta_n^{\alpha}
\right ),
\label{eq:diffusion_complete}
\end{equation}

\noindent where

\begin{equation}
\left \langle \theta ^{\alpha}  \right \rangle=\frac{1}{N}\sum_{x_{n}^{\alpha} \in G_{\alpha}} \theta_n^{\alpha}.
\end{equation}

We estimate the average value of $\dot{\theta}_n^{\alpha}$ in the layer $G_{\alpha}$, $\left \langle \dot{\theta }^{\alpha}  \right \rangle$. The result is given by

\begin{equation}
\left \langle \dot{\theta }^{\alpha}  \right \rangle=\frac{1}{N}\sum_{n=1}^{N} \dot{\theta }_n^{\alpha}= -\lambda^{12}\left ( \left \langle \theta^{\alpha} \right \rangle-\left \langle \theta^{\beta} \right \rangle \right ).
\label{eq:diffusion_prom}
\end{equation}

\noindent Note that according to Eq.~\ref{eq:diffusion_prom}, the sum of the phases of the multiplex network is constant, for $M=2$, when each layer contains a complete graph, i.e. $\left \langle \dot{\theta }^{1}  \right \rangle+\left \langle \dot{\theta }^{2}  \right \rangle=0$. Therefore,

\begin{equation}
\left \langle \theta ^{1}(t)  \right \rangle+\left \langle \theta ^{2}(t)  \right \rangle=\left \langle \theta ^{1}(0)  \right \rangle+\left \langle \theta ^{2}(0)  \right \rangle=\Gamma.
 \end{equation}

On the other hand, according to Eq.~\ref{eq:diffusion_prom}, it can be written that

\begin{equation}
\left \langle \dot{\theta }^{1}  \right \rangle-\left \langle \dot{\theta }^{2}  \right \rangle= -2\lambda^{12}\left ( \left \langle \theta^{1} \right \rangle-\left \langle \theta^{2} \right \rangle \right ).
\end{equation}

\noindent It results in

\begin{equation}
\left \langle \theta^{1}(t) \right \rangle-\left \langle \theta^{2}(t) \right \rangle=\left ( \left \langle \theta^{1}(0) \right \rangle-\left \langle \theta^{2}(0) \right \rangle \right ) e^{-2\lambda^{12}t}=\gamma e^{-2\lambda^{12}t}
\end{equation}

Hence, the evolution of the average value of $\theta^{1}$ and of the average value of $\theta^{2}$ are given by

 \begin{equation}
\left \langle \theta ^{1}(t)  \right \rangle =\frac{\gamma}{2} e^{-2\lambda^{12}t}+\frac{\Gamma}{2},
\label{eq:average_phase1}
\end{equation}

\noindent and

 \begin{equation}
\left \langle \theta ^{2}(t)  \right \rangle =-\frac{\gamma}{2} e^{-2\lambda^{12}t}+\frac{\Gamma}{2}.
\label{eq:average_phase2}
\end{equation}

By considering the series expansion

\begin{equation}
e^{i\theta_n^{\alpha}}=e^{i\left \langle \theta^{\alpha} \right \rangle}+ie^{i\left \langle \theta^{\alpha} \right \rangle}\left (\theta_n^{\alpha} -\left \langle \theta^{\alpha} \right \rangle \right )-\frac{1}{2}e^{i\left \langle \theta^{\alpha} \right \rangle}\left (\theta_n^{\alpha} -\left \langle \theta^{\alpha} \right \rangle \right )^2+\dots\quad,
\label{taylor_exp}
\end{equation}

\noindent we observe that

\begin{equation}
\sum_{x_n^{\alpha}\in G{\alpha}} e^{i\theta_n^{\alpha}}=
Ne^{i\left \langle \theta^{\alpha} \right \rangle}+
ie^{i\left \langle \theta^{\alpha} \right \rangle} \left ( \left [\sum_{x_n^{\alpha}\in G{\alpha}}\theta_n^{\alpha} \right ]
-N\left \langle \theta^{\alpha} \right \rangle \right )-\\ \nonumber
\end{equation}

\begin{equation}
-\frac{1}{2}e^{i\left \langle \theta^{\alpha} \right \rangle}
\left (
\left [ \sum_{x_n^{\alpha}\in G{\alpha}}\left ( \theta_n^{\alpha} \right )^2 \right ]
+N \left \langle \theta^{\alpha} \right \rangle ^2
-2\left \langle \theta^{\alpha} \right \rangle \sum_{x_n^{\alpha}\in G{\alpha}}\theta_n^{\alpha}
\right )+\dots\quad \approx\\ \nonumber
\end{equation}

\begin{equation}
\approx Ne^{i\left \langle \theta^{\alpha} \right \rangle}
-\frac{1}{2}e^{i\left \langle \theta^{\alpha} \right \rangle}
\left (N
\left \langle (\theta^{\alpha})^2 \right \rangle+
N \left \langle \theta^{\alpha} \right \rangle ^2
-2N \left \langle \theta^{\alpha} \right \rangle ^2
\right )
\label{tocho}
\end{equation}

We characterize the degree of synchronization of each layer $G_{\alpha}$ by means of its own order parameter, $r_{\alpha}$, expressed by

\begin{equation}
r_{\alpha}(t)e^{i\psi^{\alpha}(t) }=\frac{1}{N} \sum _{x_{n}^{\alpha}\in G_{\alpha}}e^{i\theta_{n}^{\alpha}(t) }
\rightarrow
r_{\alpha}(t)=\frac{1}{N}\left | \sum _{x_{n}^{\alpha}\in G_{\alpha}}e^{i\theta_{n}^{\alpha}(t) }\right |.
\label{eq:layer_order}
\end{equation}

\noindent Consequently, according to Eq.~\ref{tocho} and Eq.~\ref{eq:layer_order}, it is straightforward to realize that $ \psi^{\alpha}\approx \left \langle \theta^{\alpha} \right \rangle$ and

\begin{equation}
r_{\alpha}\approx1+\frac{1}{2}\left \langle \theta^{\alpha} \right \rangle ^2-\frac{1}{2}\left \langle (\theta^{\alpha})^2 \right \rangle.
\end{equation}

In case $M=2$, we obtain the following expressions for $\left \langle (\theta^{1})^2 \right \rangle$, $\left \langle (\theta^{2})^2 \right \rangle$ and $\left \langle \theta^1 \theta^2 \right \rangle$, respectively:

\begin{equation}
\left \langle (\theta^{1})^2 \right \rangle=\frac{\Gamma^2}{4} +K_1e^{-2\lambda Nt}+\frac{\gamma^2}{4}e^{-4\lambda^{12} t}-K_2e^{-2\left ( \lambda N+2\lambda^{12} \right )t }+\frac{\gamma\Gamma}{2}e^{-2\lambda^{12} t}+K_3e^{-\left ( 2\lambda N+2\lambda^{12} \right )t },
\label{theta1_cuadrado}
\end{equation}

\begin{equation}
\left \langle (\theta^{2})^2 \right \rangle=\frac{\Gamma^2}{4} +K_1e^{-2\lambda Nt}+\frac{\gamma^2}{4}e^{-4\lambda^{12} t}-K_2e^{-2\left ( \lambda N+2\lambda^{12} \right )t }-\frac{\gamma\Gamma}{2}e^{-2\lambda^{12} t}-K_3e^{-\left ( 2\lambda N+2\lambda^{12} \right )t },
\label{theta2_cuadrado}
\end{equation}

\noindent and

\begin{equation}
\left \langle \theta^1 \theta^2 \right \rangle=
\frac{\Gamma^2}{4} +K_1e^{-2\lambda Nt}-\frac{\gamma^2}{4}e^{-4\lambda^{12} t}+K_2e^{-2\left ( \lambda N+2\lambda^{12} \right )t },
\end{equation}

\noindent where $K_1$, $K_2$ and $K_3$ are constant values that depend on the initial conditions, given by

\begin{equation}
K_1=  \frac{2\left \langle \theta ^1\theta ^2  \right \rangle(0)-\Gamma ^2+ \left \langle \left ( \theta ^1 \right )^2 \right \rangle(0)+\left \langle \left ( \theta ^2 \right )^2 \right \rangle(0) }{4},
\end{equation}

\begin{equation}
K_2=\frac{2\left \langle \theta ^1\theta ^2  \right \rangle(0)+\gamma ^2- \left \langle \left ( \theta ^1 \right )^2 \right \rangle(0)-\left \langle \left ( \theta ^2 \right )^2 \right \rangle(0) }{4},
\end{equation}

\noindent and

\begin{equation}
K_3= \frac{ \left \langle \left ( \theta ^1 \right )^2 \right \rangle(0)-\left \langle \left ( \theta ^2 \right )^2 \right \rangle(0) -\gamma \Gamma }{2}.
\end{equation}

Thus, according to Eq.~\ref{eq:average_phase1}, Eq.~\ref{eq:average_phase2}, Eq.~\ref{theta1_cuadrado} and Eq.~\ref{theta2_cuadrado}, the order parameters for layers $G_1$ and $G_2$ are given by

\begin{equation}
r_1\approx 1 -\frac{1}{2}K_1e^{-2\lambda Nt}+\frac{1}{2}K_2e^{-2\left ( \lambda N+2\lambda ^{12} \right )t}-\frac{1}{2}K_3e^{-\left (  2\lambda N+2\lambda ^{12}\right )t}=\zeta -\chi,
\end{equation}

\noindent and

\begin{equation}
r_2\approx1 -\frac{1}{2}K_1e^{-2\lambda Nt}+\frac{1}{2}K_2e^{-2\left ( \lambda N+2\lambda ^{12} \right )t}+\frac{1}{2}K_3e^{-\left (  2\lambda N+2\lambda ^{12}\right )t}=\zeta+\chi,
\end{equation}

\noindent where

\begin{equation}
\zeta = \left (e^{2\lambda Nt} -\frac{1}{2} K_1-\frac{1}{2}K_2e^{-4\lambda ^{12} t} \right )e^{-2\lambda Nt},
\label{zeta}
\end{equation}

and

\begin{equation}
\chi=\frac{1}{2}K_3e^{-2\lambda ^{12}t}e^{-2\lambda Nt}.
\label{chi}
\end{equation}

Finally, the global order parameter of the multiplex network $\mathcal{M}$ (given by Eq.~\ref{eq:global_order_timescale_general}) can be approximated as

\begin{equation}
r=\sqrt{\frac{r_1^2+r_2^2+2r_1r_2\cos \left ( \Delta \right )}{4}}\approx \sqrt{ \zeta ^2 \cos^2\left (  \frac{\Delta}{2} \right )+\chi^2\sin^2\left (  \frac{\Delta}{2} \right )},
\label{final_global_order_approx}
\end{equation}

\noindent where

\begin{equation}
\Delta=\psi^1-\psi^2 \approx \left \langle \theta^{1} \right \rangle-\left \langle \theta^{2} \right \rangle=\gamma e^{-2\lambda^{12}t}.
\end{equation}



\begin{thebibliography}{99}

\bibitem{Boccaletti14}
Stefano Boccaletti, Ginestra Bianconi, Regino Criado, Charo~I Del~Genio, Jesus
  G{\'o}mez-Garde{\~n}es, Miguel Romance, Irene Sendi{\~n}a-Nadal, Zhen Wang,
  and Massimiliano Zanin.
\newblock The structure and dynamics of multilayer networks.
\newblock {\em Physics Reports}, 544(1):1--122, 2014.

\bibitem{kivela14}
M.~Kivel\"{a}, A.~Arenas, M.~Barthelemy, J.~P. Gleeson, Y.~Moreno, and M.~A.
  Porter.
\newblock Multilayer networks.
\newblock {\em J. Complex Netw.}, 2014.

\bibitem{buldyrev2010catastrophic}
Sergey~V Buldyrev, Roni Parshani, Gerald Paul, H~Eugene Stanley, and Shlomo
  Havlin.
\newblock Catastrophic cascade of failures in interdependent networks.
\newblock {\em Nature}, 464(7291):1025--1028, 2010.

\bibitem{chung1997spectral}
Fan~RK Chung.
\newblock {\em Spectral graph theory}, volume~92.
\newblock American Mathematical Soc., 1997.

\bibitem{sole13}
A.~Sol\'e-Ribalta, M.~De~Domenico, N.~E. Kouvaris, A.~D\'{\i}az-Guilera,
  S.~G\'omez, and A.~Arenas.
\newblock Spectral properties of the laplacian of multiplex networks.
\newblock {\em Phys. Rev. E}, 88:032807, Sep 2013.

\bibitem{Parshani2010}
Roni Parshani, Sergey~V. Buldyrev, and Shlomo Havlin.
\newblock Interdependent networks: Reducing the coupling strength leads to a
  change from a first to second order percolation transition.
\newblock {\em Phys. Rev. Lett.}, 105:048701, Jul 2010.

\bibitem{donges2011investigating}
Jonathan~F Donges, Hanna~CH Schultz, Norbert Marwan, Yong Zou, and J{\"u}rgen
  Kurths.
\newblock Investigating the topology of interacting networks.
\newblock {\em The European Physical Journal B}, 84(4):635--651, 2011.

\bibitem{gao2012networks}
Jianxi Gao, Sergey~V Buldyrev, H~Eugene Stanley, and Shlomo Havlin.
\newblock Networks formed from interdependent networks.
\newblock {\em Nature Physics}, 8(1):40--48, 2012.

\bibitem{PhysRevE.89.032804}
Federico Battiston, Vincenzo Nicosia, and Vito Latora.
\newblock Structural measures for multiplex networks.
\newblock {\em Phys. Rev. E}, 89:032804, Mar 2014.

\bibitem{sola2013chaos}
Luis Sol\'a, Miguel Romance, Regino Criado, Julio Flores, Alejandro
  Garc\'ia~del Amo, and Stefano Boccaletti.
\newblock Eigenvector centrality of nodes in multiplex networks.
\newblock {\em Chaos}, 23(3), 2013.

\bibitem{Domenico2014}
Manlio De~Domenico, Albert Sol\'e-Ribalta, Emanuele Cozzo, Mikko Kivel\"a,
  Yamir Moreno, Mason~A. Porter, Sergio G\'omez, and Alex Arenas.
\newblock Mathematical formulation of multilayer networks.
\newblock {\em Phys. Rev. X}, 3:041022, Dec 2013.

\bibitem{cardillo1}
Alessio Cardillo, Jes\'us G\'omez-Garde\~nes, Massimiliano Zanin, Miguel
  Romance, David Papo, Francisco Del~Pozo, and Stefano Boccaletti.
\newblock Emergence of network features from multiplexity.
\newblock {\em Sci. Rep.}, 3:1344, 2013.

\bibitem{cardillo2}
Alessio Cardillo, Massimiliano Zanin, Jes\'us G\'omez-Garde\~nes, Miguel
  Romance, Alejandro Garc\'{\i}a~del Amo, and Stefano Bocaletti.
\newblock Modeling the multi-layer nature of the european air transport
  network: Resilience and passengers re-scheduling under random failures.
\newblock {\em Eur. J. Special Topics}, 215(1):23, 2013.

\bibitem{szell}
M.~Szell, R.~Lambiotte, and S.~Thurner.
\newblock Multirelational organization of large-scale social networks in an
  online world.
\newblock {\em Proc. Natl. Acad. Sci. (USA)}, 107:13636, 2010.

\bibitem{gallotti1}
R.~Gallotti and M.~Barthelemy.
\newblock Anatomy and efficiency of urban multimodal mobility.
\newblock {\em Sci. Rep.}, 4:6911, 2014.

\bibitem{gallotti2}
R.~Gallotti, M.~Porter, and M.~Barthelemy.
\newblock Lost in transportation: Information measures and cognitive limits in
  multilayer navigation.
\newblock {\em Science Adv.}, 2:e1500445, 2016.

\bibitem{lotero}
Laura Lotero, Rafael Hurtado, Luis~Mario Flor\'{\i}a, and Jes\'us
  G\'omez-Garde\~nes.
\newblock Rich do not rise early: spatio-temporal patterns in the mobility
  networks of different socio-economic classes.
\newblock {\em Royal Soc. Open Science}, 3(10):150654, 2016.

\bibitem{radicchi}
F~Radicchi and A~Arenas.
\newblock Abrupt transition in the structural formation of interconnected
  networks.
\newblock {\em Nature Phys.}, 9:717, 2013.

\bibitem{comp}
J.~G\'omez-Garde\~{n}es, M.~De~Domenico, G.~Guti\'errez, A~Arenas, and
  S.~G\'omez.
\newblock Layer-layer competition in multiplex complex networks.
\newblock {\em Philos. Trans. Roy. Soc. A}, 373:20150117, 2015.

\bibitem{reinares}
Jes\'us G\'omez-Garde\~{n}es, Irene Reinares, Alex Arenas, and Luis~Mario
  Flor\'{\i}a.
\newblock Evolution of cooperation in multiplex networks.
\newblock {\em Sci. Rep.}, 2:620, 2012.

\bibitem{matamalas}
J~T. Matamalas, J.~Poncela-Casasnovas, and A.~Arenas.
\newblock Strategical incoherence regulates cooperation in social dilemmas on
  multiplex networks.
\newblock {\em Sci. Rep.}, 5:9519, 2015.

\bibitem{wang1}
Z.~Wang, A.~Szolnoki, and M.~Perc.
\newblock Evolution of public cooperation on interdependent networks: The
  impact of biased utility functions.
\newblock {\em EPL}, 97:48001, 2012.

\bibitem{wang2}
Z.~Wang, L.~Wang, and M.~Perc.
\newblock Degree mixing in multilayer networks impedes the evolution of
  cooperation.
\newblock {\em Phys. Rev. E}, 89:052813, 2014.

\bibitem{sorrentino12}
F.~Sorrentino.
\newblock Synchronization of hypernetworks of coupled dynamical systems.
\newblock {\em New J. Phys.}, 14:033035, 2012.

\bibitem{gambuzza15}
L~V. Gambuzza, M.~Frasca, and J.~G\'omez-Garde\~nes.
\newblock Intra-layer synchronization in multiplex networks.
\newblock {\em EPL}, 110:20010, 2015.

\bibitem{sevilla15}
R.~Sevilla-Escoboza, R.~Guti\'errez, G.~Huerta-Cuellar, S.~Boccaletti,
  J.~G\'omez-Garde\~nes, A.~Arenas, and J.~M. Buld\'u.
\newblock Enhancing the stability of the synchronization of multivariable
  coupled oscillators.
\newblock {\em Phys. Rev. E}, 92:032804, 2015.

\bibitem{genio16}
C~I. Del~Genio, J.~G\'omez-Garde\~nes, I.~Bonamassa, and S.~Boccaletti.
\newblock Synchronization in networks with multiple interaction layers.
\newblock {\em Science Adv.}, 2:e1601679, 2016.

\bibitem{kuramoto75}
Y.~Kuramoto.
\newblock Self-entrainment of a population of coupled non-linear oscillators.
\newblock In H.~Araki, editor, {\em Lecture Notes in Physics}. Springer
  Berlin/Heidelberg, 1975.

\bibitem{kuramoto84}
Y.~Kuramoto.
\newblock {\em Chemical Oscillations, Waves, and Turbulence}.
\newblock Springer Berlin Heidelberg, 1984.

\bibitem{Strogatz03}
S.~H. Strogatz.
\newblock {\em Sync: The Emerging Science of Spontaneous Order}.
\newblock Hyperion, 2003.

\bibitem{Manrubia04}
S.~C. Manrubia.
\newblock {\em Emergence of Dynamical Order: Synchronization Phenomena in
  Complex Systems}.
\newblock WorldScientific, 2004.

\bibitem{Kelly11}
D.~Kelly and G.~A. Gottwald.
\newblock On the topology of synchrony optimized networks of a kuramoto-model
  with non-identical oscillators.
\newblock {\em Chaos}, 2011.

\bibitem{Acebron05}
J.~A. Acebr\'on, L.~L. Bonilla, C.~J. P\'erez~Vicente, F.~Ritort, and
  R.~Spigler.
\newblock The kuramoto model: A simple paradigm for synchronization phenomena.
\newblock {\em Rev. Mod. Phys.}, 2005.

\bibitem{rodrigues16}
F.~A. Rodrigues, T.~K.~DM. Peron, P.~Ji, and J.~Kurths.
\newblock The kuramoto model in complex networks.
\newblock {\em Phys. Rep.}, 2016.

\bibitem{ott2008}
E.~Ott and T.~M. Antonsen.
\newblock Low dimensional behavior of large systems of globally coupled
  oscillators.
\newblock {\em Chaos}, 2008.

\bibitem{barreto2008}
E.~Barreto, B.~Hunt, E.~Ott, and P.~So.
\newblock Synchronization in networks of networks: The onset of coherent
  collective behavior in systems of interacting populations of heterogeneous
  oscillators.
\newblock {\em Phys. Rev. E}, 2008.

\bibitem{anderson2012}
D.~Anderson, A.~Tenzer, G.~Barlev, M.~Girvan, T.~M. Antonsen, and E.~Ott.
\newblock Multiscale dynamics in communities of phase oscillators.
\newblock {\em Chaos}, 2012.

\bibitem{gomez13}
Sergio Gomez, Albert Diaz-Guilera, Jesus Gomez-Garde\~{n}es, Conrad~J
  Perez-Vicente, Yamir Moreno, and Alex Arenas.
\newblock Diffusion dynamics on multiplex networks.
\newblock {\em Physical Review Letters}, 110(2):028701, 2013.

\bibitem{dedomenico15}
M.~De~Domenico, M.~A. Porter, and A.~Arenas.
\newblock Muxviz: a tool for multilayer analysis and visualization of networks.
\newblock {\em J. Complex Netwoks}, 2015.

\bibitem{serrano17}
A.~B. Serrano, J.~G\'omez-Garde\~{n}es, and R.~F.~S. Andrade.
\newblock Optimizing diffusion in multiplexes by maximizing layer
  dissimilarity.
\newblock {\em Phys. Rev. E}, 2017.

\bibitem{KMLee15}
K.M. Lee, B.~Min, and K.I. Goh.
\newblock Towards real-world complexity: an introduction to multiplex networks.
\newblock {\em Eur. Phys. J. B}, 2015.

\bibitem{dedomenico16}
M.~De~Domenico, C.~Granell, M~A. Porter, and A.~Arenas.
\newblock The physics of spreading processes in multilayer networks.
\newblock {\em Nature Phys.}, 12:901, 2016.

\bibitem{arenas2006a}
A.~Arenas, A.~D\'iaz-Guilera, and C.~J. P\'erez-Vicente.
\newblock Synchronization reveals topological scales in complex networks.
\newblock {\em Phys. Rev. Lett.}, 2006.

\bibitem{arenas2006b}
A.~Arenas, A.~D\'iaz-Guilera, and C.~J. P\'erez-Vicente.
\newblock Synchronization processes in complex networks.
\newblock {\em Physica D}, 2006.

\bibitem{Almendral07}
J.~A. Almendral and A.~D\'iaz-Guilera.
\newblock Dynamical and spectral properties of complex networks.
\newblock {\em New Journal of Physics}, 2007.

\bibitem{Grabow11}
C.~Grabow, S.~Grosskinsky, and M.~Timme.
\newblock Speed of complex network synchronization.
\newblock {\em Eur. Phys. J. B}, 2011.

\bibitem{Grabow10}
C.~Grabow, S.~Hill, S.~Grosskinsky, and M.~Timme.
\newblock Do small worlds synchronize fastest?
\newblock {\em Europhysics Letters}, 2010.

\bibitem{Son08}
S.-W. Son, H.~Jeong, and H.~Hong.
\newblock Relaxation of synchronization on complex networks.
\newblock {\em Physical Review E}, 2008.

\bibitem{avalos12}
V.~Avalos-Gayt\'an, J.~A. Almendral, D.~Papo, S.~E. Schaeffer, and
  S.~Boccaletti.
\newblock Assortative and modular networks are shaped by adaptive
  synchronization processes.
\newblock {\em Phys. Rev. E}, 2012.

\end{thebibliography}
\end{document}